\documentclass[12pt]{article}
\usepackage{amsfonts}
\usepackage{amsmath}
\usepackage{eucal}
\usepackage{amssymb}
\usepackage[perpage,symbol*]{footmisc}

\newcommand{\Rbar}{{\mbox{\rm$\mbox{I}\!\mbox{R}$}}}
\newcommand{\Hbar}{{\mbox{\rm$\mbox{I}\!\mbox{H}$}}}
\newcommand {\Cbar}
    {\mathord{\setlength{\unitlength}{1em}
     \begin{picture}(0.6,0.7)(-0.1,0)
        \put(-0.1,0){\rm C}
        \thicklines
        \put(0.2,0.05){\line(0,1){0.55}}
     \end {picture}}}
\newcommand{\beqn}{\begin{eqnarray}}
\newcommand{\eeqn}{\end{eqnarray}}
\newcommand{\arr}{\begin{array}}
\newcommand{\earr}{\end{array}}

\newcommand{\p}{\partial}

\newcommand{\uu}{\underline}

\newcommand{\BB}{{\cal B}}

\newcommand{\DD}{{\cal D}}
\newcommand{\KK}{{\cal K}}

\newcommand{\MM}{{\cal M}}

\newcommand{\LL}{{\cal L}}

\newcommand{\ZZ}{{\cal Z}}

\newcommand{\ra}{\rightarrow}

\newcommand{\ft}[2]{{\frac{#1}{#2}}}
\newcommand{\half}{{\frac{1}{2}}}
\newcommand{\fourth}{{\frac{1}{4}}}
\newcommand{\cmap}{{\bf c}-map}
\newcommand{\rmap}{{\bf r}-map}
\newcommand{\eqn}[1]{(\ref{#1})}
\newcommand{\sect}[1]{\ref{#1}}

\csname @addtoreset\endcsname{equation}{section}

\begin{document}
\begin{titlepage}
\begin{flushright}
KUL-TF-04/21\\
hep-th/0407233
\end{flushright}
\vspace{.5cm}
\begin{center}
\baselineskip=16pt {\LARGE O3/O7 Orientifold Truncations and Very Special Quaternionic-K\"ahler Geometry    \\}
\vfill
{\Large Geert Smet and Joris Van den Bergh $^\dagger$ 
  } \\
\vfill
{\small Instituut voor Theoretische Fysica, Katholieke Universiteit Leuven,\\
       Celestijnenlaan 200D B-3001 Leuven, Belgium.
 }
\end{center}
\vfill
\begin{center}
{\bf Abstract}
\end{center}
{\small We study the orientifold truncation that arises when compactifying type II string theory on Calabi-Yau
orientifolds with O3/O7-planes, in the context of supergravity. We look at the N=2 to N=1 reduction of the
hypermultiplet sector of N=2 supergravity under the truncation, for the case of very special quaternionic-K\"ahler
target space geometry. We explicitly verify the K\"ahler structure of the truncated spaces, and
we study the truncated isometry algebra. For symmetric special quaternionic spaces, we give a complete overview of the
spaces one finds after truncation. We also find new examples of `dual' K\"ahler spaces, that give rise
to flat potentials in N=1 supergravity.

 }\vspace{2mm} \vfill
 \hrule width 3.cm
 {\footnotesize \noindent
$^\dagger$  \{geert.smet, joris.vandenbergh\}@fys.kuleuven.ac.be }
\end{titlepage}

\section{Introduction}
In this paper, we study the $N=2\,\ra N=1$ reduction of the
hypermultiplet sector of $N=2$ supergravity that one gets by
performing an orientifold truncation with $O3/O7$-planes. This
truncation was discussed by Grana et al. in \cite{Louis2} and
further studied by Grimm and Louis in \cite{Louis3}. There, the
authors describe the $N=1$ low energy effective action for
compactifications of type IIB string theory on Calabi-Yau
orientifolds in the presence of background fluxes.
\\ We will study the orientifold truncation in the context of supergravi\-ty,
with the aim of describing the effect of the truncation on the
class of very special quaternionic-K\"ahler hypermultiplet moduli
spaces. These spaces arise naturally when one looks at
compac\-tifications of type IIA/IIB string theory on Calabi-Yau
spaces. For the IIA case, it is known that the low energy spectrum
contains the universal hypermultiplet (containing the dilaton, the
dual of an anti-symmetric tensor, and two scalars from the R-R
sector) and $n=h_{(1,2)}$ additional hypermultiplets. For the IIB
case, one also finds $n+1$ hypermultiplets in the low energy
spectrum, after dualizing the double-tensor multiplet and
$n=h_{(1,1)}$ additional tensor multiplets. The tree level
lagrangian for the $n+1$ hypermultiplets was calculated in
\cite{fs}. The scalar manifold spanned by the $4n+4$ scalars is
constrained by $N=2$ supergravity to be a quaternionic-K\"ahler
manifold \cite{bw}, as was explicitly verified in \cite{fs}.
Furthermore, the scalar manifold is constrained by the \cmap \ to
be of \emph{special} quaternionic type: the geometry is completely
determined by a holomorphic prepotential $F(X)$, which is
homogeneous of degree two. The very special structure arises when
one takes the large-volume limit for IIB compactifications or the
'large complex structure limit' for the IIA case (for a review,
see \cite{Hosono}): the prepotential in this limit is given by
$F=id_{ABC}\frac{X^AX^BX^C}{X^0}$ with $d_{ABC}$ a completely
symmetric tensor. Taking these limits is ne\-cessary to display
the mirror symmetry map between IIA and IIB coordinates,
as discussed in \cite{Louis}. In this paper, we will always use IIA coordinates.\\
In the supergravity context, very special quaternionic spaces arise when performing dimensional reduction from
$D=5$ to $D=3$ by applying the \rmap \ followed by the \cmap .
These maps give the following relation between very special real, K\"ahler and quaternionic-K\"ahler spaces:
\begin{equation}
\Rbar_{n-1} \stackrel{\bf r}{\longrightarrow} \Cbar_{n} \stackrel{\bf c}{\longrightarrow} \Hbar_{n+1}\ ,
\label{cmap}
\end{equation}
where $n-1$, $n$ and $n+1$ denote the real, complex and quaternionic dimensions
respectively (see \cite{symmstructure} for an exhaustive review).
\\ Compactification on a Calabi-Yau orientifold results in the $N=2 \ \ra N=1$ breaking of supersymmetry.
The effect of the orientifold truncation is to put to zero half of the scalars of the quaternionic-K\"ahler
manifold $\Hbar_{n+1}$, so that
the truncated manifold has real dimension $2n+2$. Furthermore, it has to be K\"ahler, due to the remaining $N=1$
supersymmetry.
\\ The case corresponding to the presence of only  $O3$-planes was already studied in detail by D'Auria et. al.
in \cite{Ferrara2004} and \cite{Ferrara0403} (see als \cite{stringy}, where stringy corrections were
discussed). They found that the truncated manifold factorizes into a $2n$-dimensional
K\"ahler space times the space $SU(1,1)/U(1)$. The $2n$-dimensional space can be viewed
as `dual' to the original $2n$-dimensional
K\"ahler space (denoted $\Cbar_{n}$ in \eqn{cmap}),
that is mapped by the \cmap \ into the quaternionic space.
\\ We will study the truncation of very special quaternionic-K\"ahler manifolds that corres\-ponds to the presence of both
$O3$- and $O7$-planes.
In section \sect{orientifolding} we introduce the orientifold truncation.
In section \sect{kpot} we show that the truncated
quaternionic-K\"ahler manifolds are K\"ahler, by explicitly computing the K\"ahler potential. In the following
sections we analyze the isometry algebra of the truncated manifolds:
we study in detail the case of homogeneous quaternionic spaces, and for the case of symmetric spaces, we give a complete
overview of which spaces one finds after truncation. We also extend the classification to include the symmetric special
quaternionic spaces that are not very special.
Finally, in the last section we look at the curvature tensor of the truncated manifolds and we find a new
class of K\" ahler potentials that lead to flat potentials in $N=1$ supergravity.

\section{The orientifold truncation}\label{orientifolding}

First we fix our notation: we will take the same notation as the
one used by D'Auria et al. in \cite{Ferrara2004}, except that we
rename some of the scalars to conform to the the notation used by
de Wit et. al. in \cite{symmstructure}, to which we shall refer
extensively. The $(4n+4)$-dimensional very special quaternionic
manifolds are parametrized by:
\begin{itemize}
\item the special coordinates $z^A =\half(x^A+iy^A)$, \quad $A=1,...,n$,
\item the extra coordinates $D,\sigma,A^I,B_I$, \quad $I=1,...,n+1$.
\end{itemize}
The identification with \cite{Ferrara2004} is provided by
\beqn  & &x^A \ra x^A , \quad y^A \ra y^A ,
\nonumber \\[0.3mm] & & A^I \ra \zeta^I , \quad B_I \ra \tilde \zeta_I , \cr
& &D \ra D, \quad \sigma \ra \tilde \Phi. \nonumber \eeqn The
O3/O7-orientifold truncation is described in \cite{Ferrara2004},
\cite{Louis2}: the set of special coordinates $z^A$ is separated
in two parts with opposite chirality $z^A_\pm$ $(n_+ + n_- =n)$
such that: \beqn & & y_\pm \ra \pm y_\pm , \cr
       & & x_\pm \ra \mp x_\pm ,
\eeqn
and
\beqn
& & A_\pm \ra \pm A_\mp , \quad A^0 \ra A^0 , \cr
 & & B_\pm \ra \pm B_\pm , \quad B_0 \ra -B_0 , \cr
& & D \ra D , \quad \sigma \ra -\sigma. \eeqn The case $n_{-} = 0$
(only O3-planes present) was studied in detail in
\cite{Ferrara2004, Ferrara0403}. We will extend these results to
the case $n_{-} \neq 0$. We split the index $A$ into
$A=(a,\alpha)$, with $a=1,..,n_-$ and $\alpha=1,..,n_+$. We can
now restrict to the plus-parity sector by performing the following
truncation \beqn & &x^\alpha =y^a=\sigma=0 , \cr &
&A^\alpha=B_a=B_0=0. \label{truncation} \eeqn
The remaining non-zero fields are: $\{D,x^a,y^\alpha,A^0,A^a,B_\alpha\}$. \\
For a consistent truncation, one must also demand \cite{Andrianopoli},\cite{Andrianopoli2}
\begin{equation} d_{\alpha \beta a}=d_{abc}=0.
\label{restriction} \end{equation} For later use, we introduce the
following notation \footnote{It will always be clear which
variables have a greek index, and which have a latin index, so
that there should not be any confusion between  $(\kappa
xx)_{\alpha}$ and $(\kappa yy)_\alpha$.}: \beqn (\kappa y)_{\alpha
\beta}&=&d_{\alpha \beta \gamma}y^\gamma ,\ (\kappa y)_{a b}=d_{a
b \gamma}y^\gamma , \
       (\kappa x)_{\alpha a}=d_{ab \alpha}x^b , \nonumber \\
    (\kappa yy)_\alpha&=&d_{\alpha \beta \gamma}y^\beta y^\gamma ,\
(\kappa xy)_a=d_{a b \gamma}x^b y^\gamma , \ (\kappa xx)_{\alpha}=d_{a b \alpha}x^a x^b  , \nonumber \\
      (\kappa yyy)&=&d_{\alpha \beta \gamma}y^\alpha y^\beta y^\gamma=\kappa ,\ (\kappa xxy)=d_{ab \gamma}x^a x^b y^\gamma , \
V(y)=\frac{1}{6} (\kappa yyy) .  \eeqn The $N=2$ lagrangian for
$n+1$ hypermultiplets with very special quaternionic-K\"ahler target space
is (see \cite{Louis}): \beqn \LL &=& - \half R - (\partial_{\mu}
D)^2
  - \ft14 G_{AB} \partial_{\mu} y^A \partial^{\mu} y^B
  - \ft14 G_{AB} \partial_{\mu} x^A  \partial^{\mu} x^B
  \nonumber \\ &&
  - \frac{1}{8} e^{2D} V (\partial_{\mu} A^0)^2
  - \frac{1}{2} e^{2D} V G_{AB}
  \big( x^B \partial_{\mu} A^0 - \partial_{\mu} A^B \big)
  \big( x^A \partial^{\mu} A^0 -
\partial^{\mu} A^A \big)
  \nonumber \\ &&
  - 2  e^{2D} V^{-1} G^{-1AB}
  \Big( \partial_{\mu} B_A
        + \frac{1}{8} (\kappa xx)_A \partial_{\mu} A^0
        - \frac{1}{4} (\kappa x)_{AC} \partial_{\mu} A^C \Big)\times
  \nonumber \\ && \phantom{- \half  e^{2D} V^{-1} G^{-1AB}}
  \Big( \partial^{\mu} B_B
        + \frac{1}{8} (\kappa xx)_B \partial^{\mu} A^0
        - \frac{1}{4} (\kappa x)_{BD} \partial^{\mu} A^D \Big)
  \nonumber \\ &&
  -2  e^{2D}V^{-1}
  \Big( \partial_{\mu} B_0 +  x^A \partial_{\mu} B_A
        + \frac{1}{24} (\kappa xxx) \partial_{\mu} A^0
        - \frac{1}{8} (\kappa xx)_A \partial_{\mu} A^A  \Big)^2
  \nonumber \\ &&
  - \ft14  e^{4D} \big( \partial_\mu \sigma + A^C \partial_\mu B_C
                            -B^C \partial_\mu A_C \big)^2 ,
\label{N=2lagr} \eeqn
with
\begin{equation}
 G_{AB}=-6 \left(\frac{(\kappa y)_{AB}}{(\kappa yyy)}-\ft32
\frac{(\kappa yy)_A(\kappa yy)_B
}{(dyyy)^2}\right).
\end{equation}
Performing the truncation \eqn{truncation} now gives rise to the
following lagrangian with $2n+2$ real variables (we omit the
trivially contracted index $\mu$ from here on): \beqn
{\sqrt{-g}}^{-1} {\cal{L}}&=& -(\p D)^2-\frac{1}{4} G_{\alpha
\beta} \p y^\alpha \p y^\beta - \frac{1}{4}G_{ab}\p x^a \p x^b  -
\frac{1}{8} e^{2D}V(\p A^0)^2 \cr & & -
\frac{1}{8}e^{2D}VG_{ab}(x^a\p A^0-\p A^a) (x^b\p A^0-\p A^b) \cr
& & -2e^{2D}V^{-1}(G^{-1})^{\alpha \beta} \big[\p
B_\alpha+\frac{1}{8}(\kappa xx)_\alpha \p A^0 \cr & &
-\frac{1}{4}(\kappa x)_{\alpha c} \p A^c\big]\big[\p B_\beta +
\frac{1}{8}(\kappa xx)_\beta \p A^0  - \frac{1}{4}(\kappa
x)_{\beta d} \p A^d\big], \label{lagrangian} \eeqn with truncated
elements of the metric given by \beqn G_{\alpha \beta}&=&-6
\left(\frac{(\kappa y)_{\alpha \beta}}{(\kappa yyy)}-\ft32
\frac{(\kappa yy)_\alpha(\kappa yy)_\beta }{(dyyy)^2}\right) , \cr
       G_{\alpha a} &=&0 , \cr
       G_{ab}&=&-6 \frac{(\kappa y)_{ab}}{(\kappa yyy)}.
\label{defG} \eeqn Note that the metric has no mixed components.
{}From this, and the restriction \eqn{restriction} on the
$d$-symbols, one can deduce a restriction on the matrix $C^{ABC}$
that appears in the curvature tensor (see \cite{symmstructure}).
It is defined by
\begin{equation}\label{c-up}
C^{ABC}=-27G^{AD}G^{BE}G^{CF}d_{DEF}(\kappa yyy)^{-2}
,\end{equation}
and one sees easily that
\begin{equation} C^{\alpha \beta a}=C^{abc}=0. \label{restriction2}
 \end{equation}

\section{Calculation of the K\"ahler Potential}\label{kpot}
The $N=1$ $\sigma$-model Lagrangian resulting from the
orientifolding is given by \eqn{lagrangian}. Supersymmetry requires the target space to be K\"ahler, so
the aim is to find the right coordinate transformations that allow us to write this
lagrangian in complex coordinates and to find the K\"ahler potential.
We start with the same change of variables as in \cite{Ferrara2004}
\begin{equation}
e^{2\Phi}:=V(y)e^{2D}, \qquad \chi^\alpha:=y^\alpha
e^{-\frac{\Phi}{2}},
\end{equation}

followed by
\begin{equation}
p_\alpha :=\frac{1}{2}(\kappa \chi \chi)_\alpha=\half d_{\alpha \beta
\gamma}\chi^\beta \chi^\gamma ,
\end{equation}

so that
\begin{equation}
\p \chi^\beta=((\kappa \chi)^{-1})^{\beta \alpha} \p p_\alpha.
\end{equation}

Now we define
\begin{equation} \lambda^a:=x^a e^{-\Phi}, \end{equation}
and we also perform a rescaling $A^a \rightarrow A^a \sqrt{2}$,
$A^0 \rightarrow A^0 \sqrt{2}$
and $B_\alpha \rightarrow B_\alpha /2\sqrt{2}$.
\\ This gives us the following lagrangian in the new variables:
\beqn (\sqrt{-g})^{-1} {\cal{L}} &=& -\frac{1}{4}e^\Phi G_{ab}
(\chi)[\p \lambda^a \p \lambda^b + \p A^a \p A^b]\cr &
&+\frac{1}{2} e^\Phi G_{ab}\lambda^b [-\p \Phi \p \lambda^a+e^\Phi
\p A^a \p A^0] \cr & & -\frac{1}{4}(G_{ab}+1)\lambda^a
\lambda^b[(\p \Phi)^2+ e^{2\Phi}(\p A^0)^2] \cr & &
-\frac{1}{4}V(\lambda)^{-2}{(G^{-1})}^{\alpha \beta}[\p B_\alpha
\p B_\beta + \p p_\alpha \p p_\beta + e^{2\Phi} (\kappa \lambda
\lambda)_\beta\p A^0 \p B_\alpha \cr & &-2e^\Phi(\kappa
\lambda)_{\beta c} \p B_\alpha \p A^c + \frac{1}{4}
e^{4\Phi}(\kappa \lambda \lambda)_\alpha (\kappa \lambda)_\beta
(\p A^0)^2 \cr & & e^{3\Phi} (\kappa \lambda \lambda)_\alpha
(\kappa \lambda)_{\beta c} \p A^0 \p A^c + e^{2\Phi}(\kappa
\lambda)_ {\alpha c}(\kappa \lambda)_{\beta d} \p A^c \p A^d].
\eeqn Finally, we perform the following transformation:
\begin{equation} p_\alpha = -t_\alpha + \frac{1}{2}e^{\Phi}(\kappa \lambda \lambda)_\alpha
=-t_\alpha +\frac{1}{2}e^{\Phi}d_{\alpha ab} \lambda^a \lambda^b .
\label{cotrafo}
\end{equation}
After some calculation, we find that the lagrangian can now be written in the
 fol\-lowing complex coordinates:
\beqn
& & \eta = \exp (- \Phi) + iA^o = \eta_1 + i\eta_2 , \nonumber \\
& & \epsilon_\alpha = t_\alpha + iB_\alpha , \nonumber \\
& & \rho^a=\lambda^a+iA^a . \eeqn This gives us \beqn
\sqrt{-g^{-1}}{\cal{L}} &=& -\frac{1}{4}
\big[\frac{G_{ab}}{\eta_1} +\frac{g^{\alpha \beta}}{{\eta_1}^2}
(\kappa \lambda)_{\alpha a}(\kappa \lambda)_{\beta b}\big][\p {
\rho}^a
 \p \bar \rho^b] - \frac{1}{4}g^{\alpha \beta}[\p \epsilon_\alpha \p \bar
\epsilon_\beta] \cr & &
-\frac{1}{4{\eta_1}^2}\big[1+\frac{1}{4{\eta_1}^2}g^{\alpha \beta}
(\kappa \lambda \lambda)_\alpha (\kappa \lambda
\lambda)_\beta+\frac{1} {\eta}G_{ab}\lambda^a \lambda^b\big](\p
\eta)^2 \cr & & +\frac{1}{4{\eta_1}^2}
\big[G_{ab}\lambda^b+\frac{1}{2\eta_1}g^{\alpha \beta} (\kappa
\lambda \lambda)_\alpha (\kappa \lambda)_{\beta a}\big][\p \eta \p
\bar \rho^a + \p \bar \eta \p \rho^a]
 \cr & & +  \frac{1}{4\eta_1} g^{\alpha \beta}(\kappa \lambda)_{\beta a}
[\p \epsilon_\alpha \p \bar \rho^a + \p \bar \epsilon_\alpha \p
\rho^a] \cr & & -\frac{1}{8{\eta_1}^2}g^{\alpha \beta} (\kappa
\lambda \lambda)_\beta[\p \epsilon_\alpha \p \bar \eta + \p \eta
\p \bar \epsilon_\alpha], \eeqn
with $ g^{\alpha \beta}  :=  {(G^{-1})}^{\alpha \beta}/V^2$. \\
The metric can be derived from the following K\"ahler-potential:
\begin{equation} K=-2\log V\big( \chi^\alpha(p(t_\beta,\lambda^a,\eta_1))\big) -
\log (\eta_1),
\label{kahlerpot}
\end{equation}
since we can write the Lagrangian as \beqn \sqrt{-g^{-1}}{\cal{L}}
&=& -\frac{1}{4}\big[\bar g_{a b} (\p { \rho}^a
 \p \bar \rho^b) + \bar g^{\alpha \beta}(\p \epsilon_\alpha \p \bar
\epsilon_\beta) + \bar g_{a 0}(\p \eta \p \bar \rho^a + \p \bar
\eta \p \rho^a) \cr & &  + \bar g_{00}(\p \eta)^2 + \bar
g^{\alpha}{}_{a}(\p \epsilon_\alpha \p \bar \rho^a + \p \bar
\epsilon_\alpha \p \rho^a) + \bar g^{\alpha}{}_{0}(\p
\epsilon_\alpha \p \bar \eta + \p \eta \p \bar
\epsilon_\alpha)\big], \eeqn where \beqn \bar g^{\alpha \beta} &=&
\frac{\p^2 K}{\p t_\alpha \p t_\beta} , \qquad \bar g_{0 0}=
\frac{\p^2 K}{\p \eta_1 \p \eta_1} , \qquad
\bar g_{a b}= \frac{\p^2 K}{\p \lambda^a \p \lambda^b}, \nonumber \\
\bar g_{a 0} &=& \frac{\p^2 K}{\p \lambda^a \p \eta_1} , \qquad
\bar g^{\alpha}{}_{0}= \frac{\p^2 K}{\p t_\alpha \p \eta_1} ,
\qquad \bar g^{\alpha}{}_{a}= \frac{\p^2 K}{\p t_\alpha \p
\lambda^a}. \eeqn Thus, one obtains the potential by starting with
the K\"ahler potential of \cite{Ferrara2004} and applying to it
the transformation \eqn{cotrafo}. The K\"ahler potential depends
only on the real parts of the complex variables. Contrary to the
case $n_{-} = 0$ studied in \cite{Ferrara2004}, the $(\eta,\bar
\eta)$-part does not decouple when $n_{-} \neq 0$, and therefore
does not describe a seperate $SU(1,1)/U(1)$ $\sigma$-model. The
K\"ahler space that is obtained, has complex dimension $n+1$ and
is therefore certainly distinct from the original K\"ahler space
upon which the \cmap \ is performed (which has complex dimension
$n$).
\\
It is also interesting to see whether the resulting $N=1$ target
space can give rise to no-scale models (\cite{noscale1,noscale2,
cetal}). To study this, we need the inverse metric. Defining
$(\eta^{\Lambda}) = (\eta, \epsilon_{\alpha}, \rho^a)$, we have
for the metric $\bar g_{\Lambda \bar \Sigma}$
\begin{eqnarray}
\bar g^{\alpha \beta} &=& g^{\alpha \beta}, \nonumber \\
\bar g_{0 0} &=& \frac{1}{\eta_{1}^2}(1 + \frac{1}{4 \eta_{1}^2}g^{\alpha \beta}(\kappa \lambda \lambda)_{\alpha}
(\kappa \lambda \lambda)_{\beta} + \frac{1}{\eta_{1}}G_{a b}\lambda^a \lambda^b), \nonumber \\
\bar g_{a b} &=& \frac{g^{\alpha \beta}}{\eta_{1}^2}(\kappa \lambda)_{\alpha a}(\kappa \lambda)_{\beta b} + \frac{1}{\eta_{1}}
G_{a b}, \nonumber \\
\bar g_{a 0} &=& - (\frac{1}{\eta_{1}^2} G_{a b}\lambda^b + \frac{1}{2 \eta_{1}^3}g^{\alpha \beta}(\kappa \lambda \lambda)_{\beta}
(\kappa \lambda)_{\alpha a}), \nonumber \\
\bar g^{\alpha}{}_{0} &=& \frac{1}{2 \eta_{1}^2} g^{\alpha \beta}(\kappa \lambda \lambda)_{\beta}, \nonumber \\
\bar g^{\alpha}{}_{a} &=& - \frac{g^{\alpha
\beta}}{\eta_{1}}(\kappa \lambda)_{\beta a}. \label{metric}
\end{eqnarray}

A straightforward calculation gives the following expressions for
the inverse metric $(\bar g^{-1})^{\Lambda \bar \Sigma}$

\beqn
(\bar g^{-1})_{\alpha \beta} &=& (g^{-1})_{\alpha \beta} + \frac{1}{4 \eta_{1}^2}(\kappa \lambda \lambda)_{\alpha}
(\kappa \lambda \lambda)_{\beta} + \frac{(G^{-1})^{a b}}{\eta_{1}}(\kappa \lambda)_{\alpha a} (\kappa
\lambda)_{\beta b}, \nonumber\\
(\bar g^{-1})^{0 0} &=& \eta_{1}^2 ,  \nonumber \\
(\bar g^{-1})^{a b} &=& \lambda^{a} \lambda^{b} + \eta_{1}(G^{-1})^{a b}, \nonumber \\
(\bar g^{-1})^{a 0} &=& \eta_{1} \lambda^{a}, \nonumber \\
(\bar g^{-1})^{0}{}_{\alpha} &=& \frac{1}{2}(\kappa \lambda \lambda)_{\alpha}, \nonumber \\
(\bar g^{-1})^{a}{}_{\alpha} &=& \frac{\lambda^a}{2
\eta_{1}}(\kappa \lambda \lambda )_{\alpha} + (G^{-1})^{ac}
(\kappa \lambda )_{\alpha c}. \label{invmetric} \eeqn We then find
\begin{equation}
\frac{\partial K}{\partial\Re \eta^{\Lambda}} (\bar
g^{-1})^{\Lambda \Sigma}\frac{\partial K}{\partial\Re
\eta^{\Sigma}} = 4.
\end{equation}
{}From this we see that the K\" ahler potentials obey a no scale
type condition and thus give rise to positive semi-definite
potentials \footnote{The results in this section were also found
in \cite{Louis2, Louis3}, using a different coordinate system.}.
Later on, we will show that some of the K\" ahler spaces described
in this section do factorize, namely when one performs the
truncation on the homogeneous very special quaternionic spaces. In
this case, a $SU(1,1)/U(1)$ part splits off and the remaining part
gives rise to flat potentials.
\section{Analysis of the isometries}
We analyze which isometries of the special quaternionic manifold
survive the truncation. As is discussed extensively in
\cite{symmstructure}, the isometries of a special quaternionic
manifold consist of the duality transformations inherited from the
special K\"ahler space under the \cmap, which are characterized by
parameters $\omega^i$, and in addition, $2n+4$ extra isometries
characterized by $\epsilon^0,\epsilon^+,\alpha^I, \beta_I$. There
are also hidden symmetries, which can exist when certain
conditions are satisfied by the quaternionic manifold,
characterized by $\hat \alpha^I, \hat \beta_I, \epsilon^-$. It was
proved in \cite{symmstructure} that all hidden symmetries are
realized if and only if the quaternionic manifold is symmetric.
For convenience, we repeat here the results of
\cite{symmstructure} concerning the infinitesimal transformations
(note that we have redefined $D=\half \log \phi$)
\begin{eqnarray}
\delta \phi &=&\phi \left(-\epsilon ^0+2\sigma \epsilon ^-
+\hat \alpha^I
B_I -\hat \beta_IA^I \right) ,\nonumber\\
\delta \sigma&=&\epsilon ^++\half\left(\alpha ^IB_I-\beta _IA^I\right)+
(\sigma ^2-\phi ^2)\epsilon ^-  +\sigma\left(\half
\hat \alpha^I B_I -\half\hat \beta_IA^I -\epsilon ^0\right)
+{\cal D}h ,\nonumber\\
\delta A^I&=&\alpha ^I+\sigma\hat \alpha ^I+ B^I_{\ J}(\omega)\,
A^J- D^{IJ}(\omega)\,B_J\nonumber\\
&&+\left(\epsilon ^-\sigma +\half \hat \alpha^JB_J
-\half\hat \beta_JA^J -\half\epsilon ^0\right)A^I -\partial
^I{\cal D}\left( h+\half \phi \ZZ_2 \right) ,\nonumber\\
\delta B_I&=&\beta _I+\sigma\hat\beta _I+ C_{IJ}(\omega)\,A^J-
B^J_{\;I}(\omega)\,B_J \nonumber\\
&&+ \left(\epsilon ^-\sigma +\half \hat
\alpha^JB_J -\half\hat \beta_JA^J -\half\epsilon ^0\right)B_I
+ \partial _I {\cal D}\left( h+\half \phi \ZZ_2 \right) ,\nonumber\\
\delta X^I&=& B^I_{\ J}(\omega)\,X^J +{\textstyle{\half}}i
D^{IJ}(\omega)\, F_J\nonumber\\
& &+{\cal D} \left(-\half i \bar \BB^I\,(X^J\BB_J)
+ \frac{1}{16}i(N^{-1})^{IJ}{\cal B}^K\bar F_{JKL}{\cal B}^L\,
(XN\bar X)\right) ,
\label{inftransf}
\end{eqnarray}
where
\begin{equation}
{\cal D}h =\left( \epsilon^-
+\hat\alpha{}^I\frac{\partial}{\partial A^I} +\hat\beta_I
\frac{\partial }{\partial B_I}\right) h , \label{defcalD}
\end{equation}
the function ${\cal{Z}_2}$ is given by the expression:
\begin{equation}
{\cal Z}_2\equiv {\cal B}_I\bar {\cal B}^I-2\frac{(X^I{\cal B}_I)(\bar X^J\bar
{\cal B}_J)}{XN\bar X} ,
\label{defcalZ}
\end{equation}
and the function $h$ is the real function defined by
\begin{eqnarray}
h(X,\bar X,A,B) &=&-\frac{1}{16} \left\{(\BB_I\bar
\BB^I)^2-\frac{1}{6}\left[(
F_{IJK}\bar \BB^I\bar \BB^J\bar \BB^K)(X^L\BB_L)+
h.c.\right]\right.\nonumber\\
&&\quad\left.+\frac{1}{16}(XN\bar X)
\bar \BB^I\bar \BB^JF_{IJK}N^{-1\ KL}\bar
F_{LMN}\BB^M\BB^N\right\} ,
\label{defh}
\end{eqnarray}
with
\begin{eqnarray}
\BB_I &=& B_I +{\textstyle\half}{i} F_{IJ}A^J \equiv N_{IJ}\BB^J ,
\label{defcalB}\nonumber \\
N_{IJ} &=& \frac{1}{4}(F_{IJ} + \bar F_{IJ}) ,
\end{eqnarray}
and $F(X)$ a holomorphic function which is homogeneous of second
degree in the variables $X^{I}$. The subscripts $I$, $J$, ... on
$F$ denote differentiation with respect to $X^{I}$, $X^{J}$, etc.
In these expressions, the fields $X^I$ are not independent complex
fields. The independent fields
are introduced through the ratios $z^A=X^A/X^0=\half (x^A+iy^A)$ (with $z^0:=1$). \\
In the `very special' case, $F(X)
=i\frac{d_{ABC}X^{A}X^{B}X^{C}}{X^{0}}$, and the matrix elements
$N_{IJ}$ are given by \beqn N_{00}=\frac{1}{2} i(\kappa zzz)+h.c.
\ , \quad N_{0A}=-\frac{3}{4}i(\kappa zz)_A+h.c. \ , \quad
N_{AB}=\frac{3}{2}(dy)_{AB}. \eeqn For later use, we give also the
truncated elements:
\beqn N_{\alpha \beta}&=&-\ft32(\kappa y)_{\alpha \beta} ,\ N_{ab}=-\ft32(\kappa y)_{ab}, \nonumber \\[1mm]
      N_{a\alpha}&=&N_{\alpha 0}=0 , \nonumber \\[1mm]
      N_{a0}&=&\ft34(\kappa xy)_a ,\nonumber \\[1mm]
      N_{00}&=&\frac{1}{8}(\kappa yyy)-\frac{3}{8}(\kappa xxy) .
\label{defN} \eeqn

The matrices $B,C$ and $D$ characterize the duality
transformations, and in the `very special' case, the
matrix-elements are given explicitly by
\begin{equation}
B^I_{\,J} =\left(\begin{array}{cc}
\beta & a_B \\
 b^A &\tilde B^A_{\;B} +\frac{1}{3}\beta \,\delta^A_{\,B}
\end{array}\right) ,
\ \
C_{IJ} =\left(\begin{array}{cc}
0 & 0 \\
 0 & 3 d_{ABC}\,b^C
\end{array}\right) ,
\ \
D^{IJ} =\left(\begin{array}{cc}
 0 & 0 \\
 0 & -\frac{4}{9}C^{ABC}a_C
\ \
\end{array}\right) .    \label{matrixdual}
\end{equation}
The parameters $b^A$ and $\beta$ correspond to isometries inherited from the
special K\"ahler manifold under the \cmap \ and are always present, while the
parameters $a_B$ correspond to hidden symmetries. The elements
$\tilde B^A_{\;B}$ correspond to the symmetries of $d_{ABC}$
 (see \cite{symmstructure} for details).
\\
We also list the non-zero commutators between the generators of the
isometries.
First there are those not involving the duality transformations:

\begin{equation}
\begin{array}{l}
[\underline{\epsilon}_0,\underline{\epsilon}_\pm ] =
\pm \underline{\epsilon}_\pm ,  \\[1mm]
{}[{\underline{\epsilon}}_0,\underline{\alpha}_I]=
\half \underline{\alpha}_I ,\\[1mm]
{}[\underline{\epsilon}_0,\underline{\beta}^I]=
\half \underline{\beta}^I ,\\[1mm]
{}[\underline{\epsilon}_-,\underline{\alpha}_I]=
- \underline{\hat\alpha}_I ,\\[1mm]
{}[\underline{\epsilon}_-,\underline{\beta}^I]
=- \underline{\hat\beta}{}^I ,\\[1mm]
{}[\underline{\alpha}_I,\underline{\beta}^J]
=-\delta _I^J\underline{\epsilon}_+ ,
\end{array}
\qquad
\begin{array}{l}
[\underline{\epsilon}_-,\underline{\epsilon}_+]
=2\underline{\epsilon}_0 ,\\[1mm]
{}[\underline{\epsilon}_0,\underline{\hat\alpha}_I]
=-\half\underline{\hat\alpha}_I ,\\[1mm]
{}[\underline{\epsilon}_0,\underline{\hat\beta}{}^I]
=-\half\underline{\hat\beta}{}^I ,\\[1mm]
{}[\underline{\epsilon}_+,\underline{\hat\alpha}_I]
=\underline{\alpha}_I ,\\[1mm]
{}[\underline{\epsilon}_+,\underline{\hat\beta}{}^I]
=\underline{\beta}^I , \\[1mm]
{}[\underline{\hat\alpha}_I,\underline{\hat\beta}{}^J]
=-\delta _I^J\underline{\epsilon}_- .
\end{array}
\label{commut1}
\end{equation}
The commutators involving the duality transformations are:
\begin{eqnarray}
[\alpha^I\,\underline{\alpha}_I+\beta _I\,\underline{\beta}^I,
\omega^i\,\underline{\omega}_i]
&=&{\alpha'}^I\,\underline{\alpha}_I +\beta'_I ,
\underline{\beta}^I , \nonumber\\{}
[\hat \alpha^I\,\underline{\hat\alpha}_I+\hat \beta _I \,
\underline{\hat\beta}{}^I,
\omega^i\,\underline{\omega}_i]&=&\hat {\alpha '}^I\,
\underline{\hat\alpha}_I +\hat {\beta'}_I\,\underline{\hat\beta}{}^I ,
\label{commut2}
\end{eqnarray}
where
\begin{equation}
\left( \begin{array}{c} {\alpha'}^I  \\  {\beta'}_J \end{array} \right)
= \left( \begin{array}{cc}B^I{}_K(\omega) & -D^{IL}(\omega) \\
C_{JK}(\omega) & -B^L{}_J(\omega) \end{array} \right)
\left( \begin{array}{c} \alpha^K \\  \beta_L \end{array} \right),
\label{commut3}
\end{equation}
and likewise for $\hat\alpha'$ and $\hat\beta'$. Finally
\begin{equation}
[\alpha ^I\,\underline{\alpha}_I+\beta _I\,\underline{\beta}^I,
\hat \alpha ^I\,\underline{\hat\alpha}_I+\hat \beta _I\,
\underline{\hat\beta}{}^I]=
(\alpha ^I\hat \beta _I -\hat \alpha ^I\beta _I)\,
\underline{\epsilon}_0 +\omega^i(\alpha,\beta,
\hat\alpha,\hat\beta)\, \underline{\omega}_i\ .
\label{commut4}\end{equation}
It can be shown that duality transformations corresponding to the parameters
$\omega(\alpha,\beta,\hat\alpha,\hat\beta)$ correspond to the
matrices \cite{symmstructure}
\begin{eqnarray}
B^I_{\ J}(\alpha,\beta,\hat\alpha,\hat\beta) &=&-\half(\hat\alpha
^I\beta _J+\alpha ^I\hat\beta _J)
-\partial ^I\partial _J h''(\alpha,\beta,\hat\alpha,\hat\beta) ,
\nonumber\\
 C_{IJ}(\alpha,\beta,\hat\alpha,\hat\beta)&=&-\hat\beta
_{(I}\beta _{J)} +\partial _I\partial _J h''(\alpha,\beta,
\hat\alpha,\hat\beta) ,\nonumber\\
D^{IJ}(\alpha,\beta,\hat\alpha,\hat\beta)&=&-\hat\alpha^{(I}\alpha^{J)}
+\partial ^I\partial ^J h''(\alpha,\beta,\hat\alpha,\hat\beta) ,
\label{dualitypar}
\end{eqnarray}
where
\begin{equation}
\p^I:=\frac{\p}{\p B_I} \ , \ \p_I:=\frac{\p}{\p A^I} ,
\nonumber
\end{equation}
and
\begin{equation}
h''(\alpha,\beta,\hat\alpha,\hat\beta)\equiv (\alpha \cdot
\partial +\beta \cdot\partial )  (\hat\alpha \cdot \partial
+\hat\beta \cdot\partial )\,h(X,\bar X,A,B) .
\end{equation}
{}From analyzing the infinitesimal transformations, we can now
find the remaining isometries after the truncation
\eqn{truncation}. We demand that fields that were put to zero
remain zero after performing an infinitesimal transformation. One
easily finds that the isometries corresponding to $\epsilon^0,
\alpha^a$ and $\beta_\alpha$ survive the truncation. The
truncation also restricts the possible duality transformations.
For our convenience, we start by rewriting the variation of the
fields $X^A$ as function of the $x^A$ and $y^A$. For the `very
special' case we obtain \beqn \half \delta
x^A&=&{B^A}_0+\half({B^A}_B-{B^0}_0{\delta^A}_B)x^B-
\fourth(\frac{3}{2}D^{AB}d_{BCD}+{B^0}_C{\delta^A}_D(x^Cx^D-y^Cy^D))
, \cr \half \delta y^A&=&\half({B^A}_B-{B^0}_0{\delta^A}_B)y^B-
\fourth(\frac{3}{2}D^{AB}d_{BCD}+{B^0}_C{\delta^A}_D(x^Cy^D+y^Cx^D))
, \label{trafo x,y} \eeqn disregarding for now the possible
existence of any hidden symmetries. {}From this, and the
variations of $A^I$ and $B_I$ we infer the following conditions on
the $Sp(2n+2,\Rbar)$ matrix of the duality transformations: \beqn
{B^\alpha}_0&=&b^\alpha=0 , \cr {B^\alpha}_a&=&{B^a}_\alpha=0 ,
\cr D^{\alpha \beta}&=& D^{ab}=0 . \label{dualitycond1} \eeqn
{}From the last equations, using \eqn{restriction},
\eqn{restriction2}, and \eqn{matrixdual} we can further deduce
that \beqn a_\alpha&=&0 , \cr C_{\alpha \beta}&=& C_{ab}=0 .
\label{dualitycond2} \eeqn Finally, there are the possible hidden
symmetries $\{ \hat \alpha,\hat \beta, \epsilon_- \}$. For these,
the analysis is more complicated. We start by demanding that $\DD
h=0$ after truncation, a condition that is necessary for the
variation $\delta \sigma$ to vanish. {}From this condition, a
short calculation based on results in \cite{symmstructure} for the
expression $\DD h$ reveals that the hidden isometries $\hat
\alpha^a, \hat \beta^\alpha$ and $\epsilon^-$ are certainly
broken. For the remaining isometries $\hat \alpha^\beta, \hat
\beta^a$, we must check that they satisfy the following conditions
(after truncation) \beqn \partial^\alpha \DD h &=& \partial_a \DD
h = 0 ,
   \cr   \partial^\alpha \ZZ_2 &=& \partial_a \ZZ_2 = 0 .
\eeqn After some calculation, one finds that this is indeed the
case. Finally, one can check that also the last term in the
variation of $X^I$ vanishes when calculating the variations
$\delta x^\alpha$ and $\delta y^a$ for the remaining isometries.
Crucial here is the fact that the matrix elements $N_{\alpha a}$
and $N_{\alpha 0}$ vanish identically upon truncation (see
\eqn{defN}). We conclude that the remaining isometries are
characterized by parameters $\{\alpha^0 ,\alpha^a,\hat
\alpha^\beta,\beta_\alpha,\hat \beta_0, \hat \beta_a, \epsilon_0
\}$ and by duality transformations satisfying \eqn{dualitycond1}
and \eqn{dualitycond2}. As an extra check one can verify, using
\eqn{dualitypar}, that the the duality transformations
corresponding to the hidden symmetries also satisfy
\eqn{dualitycond1} and \eqn{dualitycond2}. Note that of course
only hidden symmetries which were there in the first place, will
remain after the truncation.
\\
The same analysis can be done for the minimal coupling case, $F(X)
= X^{I}X^{J}\eta_{IJ}$, which gives rise to a class of symmetric
special quaternionic spaces \footnote{$\eta_{IJ}$ is a constant
diagonal matrix with eigenvalues $(1, -1,-1,...,-1)$.}.  In this
case, the duali\-ty transformations
 are not of the form shown in equation (\ref{matrixdual}), but obey the following equations: $B_{IJ} =
 -B_{JI}$, where
$B_{IJ} \equiv \eta_{IK}B^{K}_{J}$, and $C_{IJ} =
\frac{1}{4}\eta_{IK}D^{KL}\eta_{LJ}$. This means the group of
duality transformations is $U(n,1)$, with the generators
$B^{I}_{J}$ forming a $SO(n,1)$ subgroup.  Using the transformations
(\ref{inftransf}), which are valid for all special quaternionic
spaces, we find the following result after the truncation:
\begin{itemize}
\item broken isometries: \\$\{B^{\alpha}_{0}, B^{0}_{\alpha}, B^{a}_{\alpha}, B^{\alpha}_a,
D^{\alpha \beta}, D^{ab}, C_{\alpha \beta}, C_{ab}, \epsilon_-,
\epsilon_+ , \hat \alpha^0, \hat \alpha^a, \alpha^\beta, \hat
\beta_\alpha, \beta_0, \beta_a\}$,
\item remaining isometries: \\$\{\alpha^0 ,\alpha^a,\hat
\alpha^\beta,\beta_\alpha,\hat \beta_0, \hat \beta_a\, \epsilon_0,
B^{\alpha}_{\beta}, B^{a}_{b}, D^{\alpha a}, D^{a \alpha},
D^{\alpha 0}, D^{0 \alpha}, C_{\alpha a}, C_{a \alpha}, C_{\alpha
0}, C_{0 \alpha}\}$.
\end{itemize}
In the next section, we will take a closer look  at the isometry algebra that
arises when truncating homogeneous special quaternionic manifolds. For these
manifolds, it is known which hidden symmetries are realized.
\section{Truncation of homogeneous very special \\ quaternionic-K\"ahler
manifolds}\label{trunchvsq} The complete classification of
homogeneous very special quaternionic-K\"ahler manifolds (HVSQ
Manifolds from now on) was performed by de Wit and Van Proeyen in
\cite{classifVSQ}, and the symmetry structure of the isometries
was further studied in \cite{symmstructure}. The solutions are
denoted by $L(q, P)$, where $q$ and $P$ are integers with $q\geq
-1$ and $P\geq 0$.  For $q$ equal to a multiple of 4, there exist
additional solutions denoted by $L(4m,P,\dot P)$, with $m\geq 0$
and $P,\dot P\geq 1$. Here $L(4m,P,\dot P)=L(4m, \dot P,P)$. The
different solutions are listed in Table~\ref{classif}, where the
new spaces discovered in \cite{classifVSQ} are indicated by a
$\star$.
\begin{table}[htb]
\begin{center}
\begin{tabular}{|l|ccc|c|}\hline
$ $&Real & K\"ahler & Quaternionic  \\& (dim$_\Rbar=n-1$) &
(dim$_\Rbar=2n$) & (dim$_\Rbar=4n+4$) \\
\hline&&&\\[-3mm]
$L(-1,0)$&$SO(1,1)$&$\left[\frac{SU(1,1)}{U(1)}\right]^2$&$\frac{SO(3,4)}{(
S U ( 2 ) ) ^ 3 } $ \\[2mm]
$L(-1,P)$&$\frac{SO(P+1,1)}{SO(P+1)}$& $\star$ & $\star$ \\[2mm]
\hline&&&\\[-3mm]
$L(0,0)$&$[SO(1,1)]^2$&$\left[ \frac{SU(1,1)}{U(1)}\right] ^3$&$
\frac{SO(4,4)}{SO(4)\otimes SO(4)} $\\[2mm]
$L(0,P)$&$\frac{SO(P+1,1)}{SO(P+1)}\otimes
SO(1,1)$&$\frac{SU(1,1)}{U(1)}\otimes
\frac{SO(P+2,2)}{SO(P+2)\otimes SO(2)}$&$
\frac{SO(P+4,4)}{SO(P+4)\otimes SO(4)} $\\[2mm]
$L(0,P,\dot P)$&$Y(P,\dot P)$&$K(P,\dot P)$&$W(P,\dot P)$ \\[2mm]
$L(q,P)$&$X(P,q)$&$H(P,q)$&$V(P,q)$\\[2mm]
$L(4m,P,\dot P)$& $\star$ & $\star$& $\star$\\[2mm]
$L(1,1)$&
$\frac{S\ell(3,\Rbar)}{SO(3)}$&$\frac{Sp(6)}{U(3)
}$&$\frac{F_4}{USp(6)\otimes SU(2)}$\\[2mm]
$L(2,1)$&
$\frac{S\ell(3,\Cbar)}{SU(3)}$&$\frac{SU(3,3)}{SU(3)\otimes
SU(3)\otimes U(1)}$&$\frac{E_6}{SU(6)\otimes SU(2)}$\\[2mm]
$L(4,1)$&
$\frac{SU^*(6)}{Sp(3)}$&$\frac{SO^*(12)}{SU(6)\otimes
U(1)}$&$\frac{E_7}{\overline{SO(12)}\otimes SU(2)}$\\[2mm]
$L(8,1)$&
$\frac{E_6}{F_4}$&$\frac{E_7}{E_6\otimes
 U(1)}$&$\frac{E_8}{E_7\otimes SU(2)}$\\[2mm]
\hline \end{tabular}
\end{center}
\caption{Homogeneous special real spaces and their corresponding
K\"ahler and quaternionic spaces. The integers $P$, $\dot P$, $q$
and $m$ can take all values $\geq 1$. } \label{classif}
\end{table}
\\
The symmetric spaces are the three varieties corresponding to
$L(-1,0)$, $L(0,P)$, $L(1,1)$, $L(2,1)$, $L(4,1)$, and $L(8,1)$,
and the real spaces corresponding to $L(-1,P)$.  For those cases
the isometry group $G$ is given. For the non-symmetric spaces the
isometry group $G$ is not semisimple, the isotropy group $H$ is
always its maximal compact subgroup. \\
\\
For HVSQ manifolds the components of $d_{ABC}$ can be brought into the
cano\-nical form with non-zero components
\begin{equation}
d_{122}=1, \qquad d_{1\mu \nu }=-\delta _{\mu \nu } , \qquad
d_{2ij}=-\delta _{ij} ,\qquad d_{\mu ij}=\gamma _{\mu ij} ,
\label{dsym}
\end{equation}
where the the coordinates $h^A$ are decomposed into $h^1$, $h^2$,
$h^\mu$ and $h^i$, and the indices $\mu$ and $i$ run over $q+1$
and $r=(P+\dot P)\, {\cal D}_{q+1}$ values, respectively. Thus we
have $n= 3+q+r$.  The coefficients $\gamma_{\mu ij}$ are the
generators of a $(q\!+\!1)$-dimensional real Clifford algebra with
positive signature, denoted by ${\cal C}(q+1,0)$,
\begin{equation}
\gamma _{\mu ik}\,\gamma _{\nu kj}+\gamma _{\nu ik}\,\gamma _{\mu
kj}=2\delta _{ij}\delta _{\mu \nu } , \label{gamalgebra}
\end{equation}
and ${\cal D}_{q+1}$ denotes the dimension of an irreducible
representation of this Clifford algebra.  In the case $q \neq 0$
mod~4, the irreducible representations for a given $q$ are unique
and thus the gamma matrices are unique once we specify the number
of irreducible representations, denoted by $P$.  For $q=0$ mod~4 the
representations $\gamma _\mu $ and $-\gamma _\mu $ are not
equivalent, and a reducible representation is characterized by the
multiplicity of each of these representations, $P$ and $\dot P$.
Hence $r=(P+\dot P)\, {\cal D}_{q+1}$ in general.  The symmetric
spaces are characterized by $\Gamma_{ijkl}=0$, with
$\Gamma_{ijkl}$ defined by
\begin{equation}
\Gamma _{ijkl}\equiv \frac{3}{8} \left[ \gamma _{\mu (ij}\,\gamma
_{kl)\mu }-\delta _{(ij}\, \delta_{kl)}\right] .
\label{defGamma}
\end{equation}
\\
For calculational purposes it's sometimes convenient to combine the
indices 2 and $\mu$ in on index $M$.  We can then rewrite the
relations in section \ref{trunchvsq}.  The non-zero components of
the tensor $d_{ABC}$ are
\begin{equation}
d_{1MN}= -\eta _{MN} , \qquad d_{Mij}=\gamma _{Mij} .
\label{dSOq+11}
\end{equation}
The $\gamma _{Mij}$ obey
\begin{equation}
\gamma _{M ik}\,\gamma _{N}{}^{kj}+\gamma _{N ik}\,\gamma _{M}{}^{kj} = 2\delta _{i}^{j}\eta _{M N } ,
\end{equation}
with $\eta = (-1,1,1,...,1)$, $\gamma_{\mu\,ij}=\gamma_\mu{}^{ij}$
and $ \gamma_2{}^{ij}= -\gamma_{2\, ij}= \delta_{ij}$. The
symmetric spaces are then characterized by $\Gamma
_{ijkl}=\frac{3}{8}\left[ \gamma _{M (ij}\,\gamma _{kl) N}
\eta^{MN}\right] = 0 $.
\\
For the orientifold truncation, we take the index $\alpha$ of the
previous sections to run over $1$ and $M$ (this is necessary,
considering the restriction \eqn{restriction} on the $d$-symbols),
while the index $a$ is identified with the index $i$ \footnote{The case $(\alpha) = (1,M,i)$ was studied in
\cite{Ferrara2004, Ferrara0403}.} (this can be
generalized in some cases, as we will show in section \ref{generalisation}). The
symmetries of the $d$-symbols are encoded in the matrix ${\tilde
{B^A}_B}$, which appears in the duality transformations and has
the following structure (see \cite{symmstructure})
\begin{equation}
\tilde B^A{}_{\!B}=\left( \begin{array}{ccc} -2\lambda & 0& 2\xi _j
       \\     0&\lambda\,\delta ^M_N+A^M{}_{N} & -\zeta^k\,\gamma^M{}_{kj}
        \\    -\zeta^i & \gamma _N{}^{ik}\xi_k & -\half \lambda
    \delta^i_j +\ft14 A^{PQ}\left( \gamma _{PQ}\right)^i{}_{j}
    +S^i{}_{j} \end{array} \right) .
\label{d-symmetries}
\end{equation}
The matrices ${A^M}_N$ generate infinitesimal
$SO(q+1,1)$-symmetries and ${S^i}_j$ generate the group ${\cal
S}_q(P,\dot P)$ which is the centralizer of the Clifford algebra
representation \cite{symmstructure}. Now using \eqn{dualitycond1},
we can see that all isometries will survive the orientifold
truncation, except the ones with parameters $\xi_j$ and $\zeta^j$,
which are broken. The full set of isometries of the HVSQ manifold
consists of the above isometries coming from the symmetries of the
$d$-symbols, plus the extra symmetries one gets from performing
the \rmap \ and \cmap. For \emph{non-symmetric} HVSQ manifolds,
the following hidden symmetries are realized:
$\{\uu{a}^M,\underline{\hat \alpha}{}_1,\underline{\hat
\beta}{}^M, \underline{\hat \beta}{}^0\}$. The analysis of the
various commutation relations and the isometry algebras that
arise, is performed in \cite{symmstructure}, resulting in the
classification given in table~\ref{classif}. We will not repeat
all the results here, and we refer to \cite{symmstructure} for the
details. Our aim is to study what happens to these manifolds upon
performing the orientifold truncation. The important results are
that the resulting K\"ahler manifolds are also homogenous
manifolds that have one half of the dimension of the HVSQ
manifold. Furthermore, they are always product manifolds of the
form $\MM \times SU(1,1)/U(1)$, with $\MM$ having the same
dimension as the original special K\"ahler space that is the
target of the \cmap .

\subsection{Homogenity}
We have to show that the remaining isometries work transitively on
the truncated mani\-fold. It is obvious that this is so for the
remaining shifts and scale transformation. To study the effect of
the remaining duality transformations, we write down the truncated
version of \eqn{trafo x,y}
\beqn
\delta x^a &=& b^a+{B^a}_b x^b -
\beta x^a-\frac{3}{4}D^{a \alpha}(d_{\alpha cd}x^c x^d+d_{\alpha
\beta \gamma}y^\beta y^\gamma)-\half a_c x^c x^a  , \cr
 \delta y^\alpha &=& {B^\alpha}_\beta y^\beta -
 \beta y^\alpha - \frac{3}{2} D^{\alpha b}d_{bc \gamma}x^c
y^\gamma-\half a_c x^c y^\alpha .
\eeqn
Looking at the form of the matrix ${\tilde B^A}_B$ in
\eqn{d-symmetries}, it is now clear that one can connect any two
points in the truncated mani\-fold by performing shifts $b^a$ (for
the co\"ordinates $x^a$), rotations ${A^M}_N$ and scale
transformations $\beta$ (for the co\"ordinates $y^\alpha$).
Thus, the truncated manifold is homogeneous.
To find the elements of the isotropy group, we can choose a point
with $A^0=A^a=B_\alpha =x^a=y^\mu=0$ (i.e. only $y^1,y^2$
different from $0$). This point is left invariant by generators of
${\cal S}_q(P,\dot P)$ and $SO(q+1)$, and by the generators of the
hidden symmetries.

\subsection{Splitting of an $SU(1,1)/U(1)$-part}\label{splitting}
We will show that there exists an $SU(1,1)$- subalgebra of the isometry
algebra, of which the three generators commute with all the other generators.
First, we give a list of the isometries of the HVSQ spaces that survive the
truncation:
\\ $\{\uu{\lambda},{\uu{A}^M}_N,{\uu{S}^i}_j,\uu{\beta},\uu{\epsilon}_0,
 \uu{\beta}^M,\uu{\alpha}_i,\uu{b}_i,\uu{\hat \alpha}_1,\uu{\hat \beta}^0
,\uu{\alpha}_0,\uu{\beta}^1\}$ . \\
Next, we write down the commutators between the generators
$\underline{\alpha}$,$\underline{\beta}$, $\uu{a}$ and $\uu{b}$
 which were not given explicitly
in \cite{symmstructure}. Using \eqn{commut4}, one finds the following relations
\beqn
&& \arr{ll}
[\uu{\alpha}_0,\uu {\hat \alpha}_\beta]=0 , &
[\uu{\hat \beta}^0,\uu{\alpha}_a]=\uu{b}_a ,
\\{} [\uu {\alpha}_0, \uu {\hat \beta}^b]=-\uu {a}^b , &
[\uu{\hat \beta}^0,\uu{\beta}^\alpha]=0 ,
\earr
 \cr
&& \arr{l}
[\uu {\alpha}_0, \uu {\hat \beta}^0]=\uu {\epsilon}_0-\frac{3}{2}
\uu{\beta},
\\{}
[\uu{\alpha}_a, \uu{\hat \alpha}_\beta]=3d_{\beta ab} \uu{a}^b ,
\\{}
[\uu{\beta }^\alpha, \uu{\hat \beta}^a]=-\frac{4}{9}C^{\alpha ab} \uu{b}_b ,
\\{}
[\uu{\beta}^1, \uu{\hat
\alpha}_1]=-\uu{\epsilon}_0+\frac{2}{3}\uu{\lambda}-\half
\uu{\beta} ,
\\{}
[\uu{\beta}^2, \uu{\hat
\alpha}_2]=-\uu{\epsilon}_0-\frac{1}{3}\uu{\lambda}-\half
\uu{\beta} ,
\\{}
[\uu{\beta}^\mu, \uu{\hat
\alpha}_\nu]=-\uu{A}_{\mu \nu}+\delta_{\mu \nu}(-\uu{\epsilon}_0-\frac{1}{3}\uu{\lambda}-\half
\uu{\beta}) ,
\earr
\cr
&& \arr{ll}
[\uu{a}^b,\uu{\hat \beta}^0]= \hat \beta^b , &
[\uu{b}_a,\uu{\alpha}_0]=-\uu{\alpha}_a , \\{}
[\uu{a}^b,\uu{\hat \alpha}_\beta]=0 , &
[\uu{b}_a,\uu{\alpha}_b]=-3d_{\alpha ab}\uu{\beta}^\alpha , \\{}
[\uu{a}^b,\uu{\hat \beta}^a]=-\frac{4}{9}C^{\beta ab}\uu{\alpha}_\beta , &
[\uu{b}_a,\uu{\beta}^\alpha]=0 .
\earr
\label{commut5}
\eeqn
When one now considers the explicit form of the $d$-symbols and
the tensor ${C^{ABC}}$ (\cite{symmstructure}), one finds that the
set of generators $\{ \uu{\hat \alpha}_1, \uu{\beta}^1
,\uu{\epsilon} \equiv -\uu{\epsilon}_0+\frac{2}{3}\uu{\lambda}-\half
\uu{\beta}\}$ generates an $SU(1,1)$-subalgebra, and furthermore commutes
with all other generators. The K\"ahler space resul\-ting from the truncation
will thus contain a $SU(1,1)/U(1)$-part which splits off from the manifold,
with  the $U(1)$ subgroup generated by the compact generator $\hat {\alpha}_1$.
This result will be confirmed when we analyze the metric and curvature tensor
later on.

\section{Truncation of symmetric special quaternionic-\\ K\"ahler
 manifolds}\label{truncssqm}
\subsection{SVSQ manifolds: the generic truncation}\label{generic}
In the previous section, we showed that a truncated HVSQ manifold takes the
form $\MM \times SU(1,1)/U(1)$. It is therefore interesting to treat the
manifold $\MM$ as a `dual' K\"ahler manifold in the sense of
\cite{Ferrara2004}, and to look closer at its properties. We shall start by loo- \\ king at the subclass of
\emph{symmetric} very special quaternionic-K\"ahler manifolds
(SVSQ manifolds).
These manifolds take the form of coset spaces $G/H$, where the isometry group
$G$ is a semisimple Lie-group, and the isotropy group $H$ is its maximal
compact subgroup.
The SVSQ  manifolds are the ones denoted by $L(-1,0)$, $L(0,P)$, $L(1,1)$,
$L(2,1)$, $L(4,1)$, and $L(8,1)$ in table~\eqn{classif}.
For these manifolds, all hidden symmetries are realized. The full
set of isometries and
the commutators are given in table~\ref{rootsVSQ}, which we have taken from
\cite{symmstructure}, completed with the additional hidden symmetries.
\begin{table}[htf]
\caption{Roots of the isometries of the symmetric
 very special quaternionic spaces: \quad  $so(q+2,2)$ is generated by
$\{{\uu{A}^M}_N,\uu{a}^M,\uu{b_M},
\uu{\beta}-\frac{1}{3} \uu{\lambda} \}$ and $\lambda' \equiv \ft23 \lambda
 + \beta $}.
\begin{center}\begin{tabular}{||l|ccrr|rr||}\hline
generator &${\cal S}_q$& $so(q+2,2)$ & $\underline{\lambda}'$  &
$\underline{\epsilon}{}_0$ &$\underline{\lambda}'-2
\underline{\epsilon}{}_0$&
$\underline{\lambda}'+2\underline{\epsilon}{}_0$
\\[2mm] \hline
$\underline{\epsilon}{}_+$&0&0&0&1&$-2$&2\\[2mm]
$(\underline{\alpha}{}_1,\underline{\beta }^M,\underline{\beta }^0)$
&0& $v$&1&$\ft12 $&0&2 \\[2mm]
$\underline{b}_1$&0& 0 & 2& 0&2&2\\[2mm]
$(\underline{\alpha}{}_i,\underline{\beta}{}^i)$ & $v$&$s$ &0&$\ft12$&$-1$&1
\\[2mm]
$(\underline{\xi}^i,\underline{b}{}_i)$& $v$&$\bar s$ &1&0&1&1\\[2mm]
$(\underline{\hat \alpha}{}_1,\underline{\hat \beta}{}^M,\underline{\hat
\beta}{}^0)$&0& $v$&1&$-\ft12$& 2&0 \\[2mm]
$(\underline{\alpha }_0,\underline{\alpha }_M,\underline{\beta }^1)$
&0& $v$&$-1$&$\ft12$&$- 2$&0 \\[2mm]
 \hline
$\underline{\epsilon}{}_-$&0&0&0&$-1$&$2$&$-2$\\[2mm]
$(\underline{\hat \alpha}{}_0,\underline{\hat \alpha }_M,
\underline{\hat \beta }^1)$
&0& $v$&$-1$&$-\ft12 $&0&$-2$ \\[2mm]
$\underline{a}^1$&0& 0 & $-2$& 0&$-2$&$-2$\\[2mm]
$(\underline{\hat \alpha}{}_i,\underline{\hat \beta}{}^i)$
& $v$&$\bar s$ &0&$-\ft12$&$1$&$-1$ \\[2mm]
$(\underline{\zeta}_i,\underline{a}{}^i)$& $v$&$s$ &$-1$&0&$-1$&$-1$\\[2mm]
\hline
\end{tabular}\end{center}
\label{rootsVSQ}
\end{table}
\\
The isometry algebra can be decomposed as follows (see \cite{symmstructure})
\begin{eqnarray}
{\cal V}&=& {\cal V}_0 +{\cal V}_{1} + {\cal V}_2 + {\cal V}_{-1} + {\cal V}_{-2} ,\nonumber\\
{\cal V}_0 &=&\underline{\epsilon }'\oplus so(q+3,3)\oplus {\cal
S}_q(P,\dot P) , \nonumber\\
{\cal V}_1 &=& (\underline{\xi}^i,\underline{b}_i)
\oplus(\underline{\alpha}{}_i,\underline{\beta}{}^i) =  (1,s,v) ,
\nonumber\\
{\cal V}_2 &=& \underline{\epsilon}{}_+\oplus
(\underline{\alpha}{}_1,\underline{\beta}{}^M,
\underline{\beta}{}^0) \oplus \underline{b}{}_1 = (2,v,0) , \nonumber \\
{\cal V}_{-1} &=& (\underline{a}^i,\underline{\zeta}_i)
\oplus(\underline{\hat\alpha}{}_i,\underline{\hat\beta}{}^i) =
(-1,s,v) ,\nonumber\\
{\cal V}_{-2} &=& \underline{\epsilon}{}_-\oplus
(\underline{\hat\alpha}{}_0,\underline{\hat \alpha}{}_M,
\underline{\hat\beta}{}^1) \oplus \underline{a}{}^1 = (-2,v,0) ,
\label{iso}
\end{eqnarray}
where the representations of ${\cal V}_{1}$,${\cal V}_2$,
${\cal V}_{-1}$ and ${\cal V}_{-2}$ are indicated according to the three
subalgebras of ${\cal V}_{0}$ and
$\underline{\epsilon}' \equiv 2\uu{\epsilon}_{0} + \uu{\beta} + \frac{2}{3}\uu{\lambda}$.
\\
After the orientifold projection, the following set of isometries
will survive: \\
$\{\uu{\lambda},{\uu{A}^M}_N,{\uu{S}^i}_j,\uu{\beta},\uu{\epsilon}_0,
 \uu{\beta}^M,\uu{\alpha}_i,\uu{b}_i,\uu{\hat \alpha}_1,\uu{\hat \beta}^0
,\uu{\alpha}_0,\uu{\beta}^1,\uu{a}^i,\uu{\hat \beta}^i,
\uu{\hat \alpha}_M\}$ . \\
Counting the number of compact and non-compact generators gives us:
\begin{itemize}
\item dim$\,G=2n+6+\half q(q+3)+2r+dim\,{\cal S}_q$ \ remaining isometries ,
\item dim$\,H=n+1+\half q(q+1)+r+dim\,{\cal S}_q$ \ remaining compact isometries ,
\end{itemize}
so that the dimension of the truncated manifolds is given by
\begin{equation}
dim\,\MM_{trunc}=dim\,G-dim\,H=n+r+q+5=2n+2 .
\end{equation}
This is consistent with the result $\MM_{trunc}=\KK_{dual} \times
SU(1,1)/U(1)$. Also, looking at the root diagrams presented in
\cite{symmstructure} and \cite{DWVP} reveals the following fact:
if a root corresponds to an isometry that is projected out, the
opposite root also belongs to a broken isometry. In other words,
opposing roots are always projected out together.
 \\ After performing the truncation on the isometry algebra
\eqn{iso} and throwing away the $SU(1,1)/U(1)$ part that
splits off, we find the following structure for the isometry
algebra of $\KK_{dual}$ \beqn &&{\cal W}={\cal W}_0 +  {\cal W}_1
+  {\cal W}_2 + {\cal W}_{-1} +  {\cal W}_{-2} ,
\nonumber\\
&&{\cal W}_0= \underline{\epsilon}' \oplus
(\uu{\epsilon}_{0}-\frac{3}{2}\uu{\beta},\uu{\alpha}_{0},\uu{\hat
\beta}_{0}) \oplus so(q+1,1)\oplus {\cal
S}_q(P,\dot P) ,\nonumber\\
&&{\cal W}_1= \underline{b}^i \oplus \underline{\alpha}_i
=(1,s,s,v)\
,\nonumber\\
&&{\cal W}_2= \underline{\beta}^M =(2,0,v,0) , \nonumber \\
&&{\cal W}_{-1}= \underline{a}_i \oplus \underline{\hat \beta}^i =(-1,s,s,v) ,\nonumber\\
&&{\cal W}_{-2}= \underline{\hat \alpha}_M=(-2,0,v,0) . \eeqn
{}From this, we see that the generators $\uu{\epsilon}'$,
$\uu{\beta}^{M}$, $\uu{\hat \alpha}_{M}$, together with the
generators $\uu{A}_{MN}$ of $so(q+1,1)$ form the algebra
$so(q+2,2)$. We can then decompose the isometry algebra with
respect to the grading under
$\uu{\epsilon}_{0}-\frac{3}{2}\uu{\beta}$ \beqn &&{\cal W}'={\cal
W}'_0 +  {\cal W}'_1 +  {\cal W}'_2 + {\cal W}'_{-1} +  {\cal
W}'_{-2} ,
\nonumber\\
&&{\cal W}'_0= \uu{\epsilon}_{0}-\frac{3}{2}\uu{\beta} \oplus
so(q+2,2) \oplus {\cal
S}_q(P,\dot P) ,\nonumber\\
&&{\cal W}'_1= \underline{a}^i \oplus \underline{\alpha}_i =(1,s,v)\
,\nonumber\\
&&{\cal W}'_2= \underline{\alpha}_0 =(2,0,0) , \nonumber \\
&&{\cal W}'_{-1}= \underline{b}_i \oplus \underline{\hat \beta}^i =(-1,s,v) ,\nonumber\\
&&{\cal W}'_{-2}= \underline{\hat \beta}_0=(-2,0,0) .
\eeqn
This has exactly the same structure as the isometry algebra of the symmetric very special K\"ahler spaces (see
\cite{symmstructure}).  The isotropy group is again the maximal compact subgroup of the isometry group.
Another way to get this result, is by looking at all possible semi-simple subgroups of the isometry groups of the
symmetric very special quaternionic manifolds and imposing the constraints that these subgroups have the right dimension
and the right number of non-compact generators.  One then always finds a unique possibility.
\\ {}From these results, we can conclude that the dual K\"ahler spaces are the same cosets as the original
special K\"ahler spaces. Note that this result does not
necessarily imply that the dual space has the same curvature. We
will investigate this in the next section.
\subsection{A generalisation for the case $L(0,P)$}\label{generalisation}
When $P > 1$, we can split the index $i$ in two parts, $(i) =
(i_{+}, i_{-})$ where $i_{+}$ runs over $P_{+}{\cal D}_{q+1}$
values, $i_{-}$ runs over $P_{-}{\cal D}_{q+1}$ values, $P = P_{+}
+ P_{-}$, and in such a way that $\gamma_{\mu i_{+} i_{-}} = 0$.
We can take the $\alpha$-index of the previous sections to run
over $1,2,\mu$ and $i_{+}$, and identify the $a$-index with
$i_{-}$. Therefore, from now on, we will write the $i_{+}$-index
as $\tilde{M}$ and the $i_{-}$-index as $a$. We can then start
with the isometry algebra for the very special quaternionic spaces
given in (\ref{iso}), specialised to the case $q=\dot P=0$, and
with $(i)=(\tilde{M},a)$. After the truncation, the $SU(1,1)$
generated by $\{ \uu{\hat \alpha}_1, \uu{\beta}^1 ,\uu{\epsilon}
\equiv -\uu{\epsilon}_0+\frac{2}{3}\uu{\lambda}-\half
\uu{\beta}\}$, does not split off anymore if $P_{+} \neq 0$. We
find the following structure for the truncated isometry algebra
\beqn
&&{\cal W}={\cal W}_0 +  {\cal W}_{+} +  {\cal W}_{-} , \nonumber\\
&&{\cal W}_0= SO(2,2)_{+} \oplus SO(P_{+}) \oplus SO(2,2)_{-} \oplus SO(P_{-}) , \nonumber\\
&&{\cal W}_{+}= (\underline{\xi}^{\tilde{M}},\underline{\hat
\alpha}_{\tilde{M}}, \underline{\beta}^{\tilde{M}},
\underline{\zeta}_{\tilde{M}}) = (v,v,0,0) ,\nonumber\\
&&{\cal W}_{-}= (\underline{a}^b,\underline{\alpha}_b,
\underline{b}_a, \underline{\hat \beta}^a) = (0,0,v,v) , \eeqn where
\begin{itemize}
\item $SO(2,2)_{+}$ is generated by $\{ \uu{\hat \alpha}_1, \uu{\beta}^1, \uu{\epsilon}, \uu{\beta}^2 +\uu{\beta}^3, \uu{\alpha}^2 +\uu{\alpha}^3,
-\uu{\epsilon}-2\uu{A}^2_3\}$ ,
\item
$SO(2,2)_{-}$ by $\{ \uu{\hat
\beta}^0, \uu{\alpha}_0, \uu{\epsilon}_0- \frac{3}{2} \uu{\beta},
\uu{\beta}^2 -\uu{\beta}^3, \uu{\alpha}^2
-\uu{\alpha}^3,-\uu{\epsilon}+2\uu{A}^2_3\}$ ,
\item $SO(P_{+})$ by
$S^{\tilde{M}}_{\tilde{N}}$ and $SO(P_{-})$ by $S^{a}_{b}$.
\end{itemize}
It is also easy to see that the generators in $W_{+}$ commute with
those in $W_{-}$, since their commutators will be proportional to
either $\delta^{a\tilde{M}}$, $\gamma^{\mu a\tilde{M}}$ or
$S^{a\tilde{M}}$ and those are all zero. Thus, we find that the
isometry group surviving the truncation is $SO(P_{+}+2,2) \times
SO(P_{-}+2,2)$. It is again straightforward to show that these
isometries work transitively on the manifold, so that the
resulting K\" ahler space is\footnote{For $P_{+} = 0$, which was
discussed in section \ref{generic}, or for $P_{-} = 0$, which was
discussed in \cite{Ferrara2004, Ferrara0403} this reduces to
$[SU(1,1)/U(1)]^2\times [SO(P+2,2)/(SO(P+2)\times SO(2))]$.}
\begin{equation}
\frac{SO(P_{+}+2,2)}{SO(P_{+}+2)\times SO(2)}\times
\frac{SO(P_{-}+2,2)}{SO(P_{-}+2)\times SO(2)}.
\end{equation}
\subsection{The other SSQ manifolds}
Finally, we can try to extend our classification
by analyzing the effect of the truncation on symmetric
special quaternionic manifolds that are not very special.
There are three cases to consider: (see \cite{DWVP} for an overview)
\begin{itemize}
\item The space $U(1,2)/(U(1)\times
U(2))$, which one gets by acting with the \cmap \ on the empty
special K\"ahler space.
\item The space $G_2/(SU(2)\times SU(2))$, resulting from the \cmap \
on the special K\" ahler manifold $SU(1,1)/U(1)$ that is
characterized by the prepotential $F = i(X^1)^3/X^0$.
\item The spaces denoted by $L(-2,P)$ in \cite{DWVP},
resulting from the \cmap \ on \\ $U(P+1,1)/(U(P+1)\times U(1))$:
K\"ahler spaces characterized by a quadratic pre\-potential $F(X)=
X^I\eta_{IJ}X^J$. Because of the quadratic prepotential, the truncation conditions
from \cite{Andrianopoli2} are trivially satisfied, so that no extra conditions need to be imposed.
This means that different truncations are possible,
depending on how we split the index $I$ into indices $(0,a,\alpha)$.
\end{itemize}
The analysis of the duality symmetries is performed in
\cite{DWVP2} and \cite {DWVP3}, to which we refer for details.
These are the results for the truncated manifolds \footnote{As
before the remaining isometries work transitively, so that we
again find homogeneous K\" ahler submanifolds of the corresponding
quaternionic spaces.}
\begin{itemize}
\item
In the first case, the coset $SU(1,1)/U(1)$, the set of three surviving
isometries being $\{ \uu{\alpha}_0, \uu{\hat \beta}^0
, \uu{\epsilon}_0\}$.
\item
In the second case, the coset $(SU(1,1)/U(1))^2$, with the two
commuting $SU(1,1)$-subalgebras generated by $\{ \uu{\alpha}_0,
\uu{\hat \beta}^0 , \uu{\epsilon}_0-\frac{3}{2} \uu{\beta}\}$ and
$\{ \uu{\hat \alpha}_1, \uu{\beta}^1, -\uu{\epsilon}_0-\half
\uu{\beta}\}$. (Although the space is usually not considered to be
`very special', the duality transformations are still of the form
(\ref{matrixdual}).)
\item
In the third case, the structure of the isometry algebra before the truncation is
\begin{eqnarray}
{\cal V}&=& {\cal V}_0 +{\cal V}_{\frac{1}{2}} + {\cal V}_1 + {\cal V}_{-\frac{1}{2}} + {\cal V}_{-1} ,\nonumber\\
{\cal V}_0 &=&\underline{\epsilon}{}_{0}\oplus U(P+1,1) , \nonumber\\
{\cal V}_{\frac{1}{2}} &=& (\underline{\alpha}{}_I,\underline{\beta}{}^I) = (\frac{1}{2}, P+2) \oplus
(\frac{1}{2}, \overline{P+2}) , \nonumber\\
{\cal V}_1 &=& \epsilon^{+} , \nonumber \\
{\cal V}_{-\frac{1}{2}} &=& (\underline{\hat \alpha}{}_I,\underline{\hat \beta}{}^I) = (-\frac{1}{2}, P+2) \oplus
(-\frac{1}{2}, \overline{P+2}) , \nonumber\\
{\cal V}_{-1} &=& \epsilon^{-} ,
\end{eqnarray}
forming $SU(P+2,2)$.
For all possible truncations, we find the same result:
a total number of $\half(P+4)(P+3)$ isometries is preserved,
of which $2(P+2)$ are non-compact.  The truncated isometry algebra has the following structure:
\begin{eqnarray}
{\cal V}&=& {\cal V}_0 +{\cal V}_{\frac{1}{2}} + {\cal V}_{-\frac{1}{2}} ,\nonumber\\
{\cal V}_0 &=&\underline{\epsilon}{}_{0}\oplus SO(P+1,1) , \nonumber\\
{\cal V}_{\frac{1}{2}} &=& (\underline{\alpha}{}_0,\underline{\alpha}{}_a,\underline{\beta}{}^{\alpha}) = (\frac{1}{2}, v) , \nonumber\\
{\cal V}_{-\frac{1}{2}} &=& (\underline{\hat \alpha}{}_0,\underline{\hat \alpha}{}_a ,\underline{\hat
\beta}{}^{\alpha}) = (-\frac{1}{2}, v) ,
\end{eqnarray}
forming $SO(P+2,2)$. {}From this we conclude that the truncated
spaces consist of the class $SO(P+2,2)/(SO(P+2)\times SO(2))$.
These are symmetric K\"ahler spaces that are not special (see
\cite{CremmerVP} for the classification). Note that no
$SU(1,1)$-part splits off.
\end{itemize}

The results for the truncation of all SSQ manifolds are summarized in
table~\ref{results}.
\begin{table}[htb]
\caption{Truncation of symmetric special quaternionic manifolds}
\begin{center}
\begin{tabular}{|l|c|c|}
    \hline&&\\[-3mm]
  &  ${\cal M}_Q$ (dim$_\Rbar=4n+4$) &
    ${\cal M}_{KH}$ (dim$_\Rbar=2n+2$) \\
    \hline&&\\[-3mm]
 $SG_4$ & $\frac{U(2,1)}{U(2)\times U(1)}$ &  $\frac{SU(1,1)}{U(1)}$ \\[2mm]
 $L(-2,P)$ & $\frac{SU(P+2,2)}{SU(P+2)\times SU(2)\times U(1)} $
& $\frac{SO(P+2,2)}{SO(P+2) \times SO(2)}$ \\[2mm]
 $L(-1,0)$ & $\frac{SO(3,4)}{(SU(2)^3)}$ & $[\frac{SU(1,1)}{U(1)}]^3$ \\[2mm]
$L(0,0)$ & $\frac{SO(4,4)}{SO(4)\times SO(4)}$ & $[\frac{SU(1,1)}{U(1)}]^4$
\\[2mm] $L(0,P)$&    $\frac{SO(P+4,4)}{SO(P+4)\times SO(4)} $ &
    $\frac{SO(P_{+}+2,2)}{SO(P_{+}+2)\times
SO(2)}\times \frac{SO(P_{-}+2,2)}{SO(P_{-}+2)\times SO(2)}$  \\[2mm]
$SG_5$   &  $\frac{G_{2(2)}}{SO(4)} $& $[\frac{SU(1,1)}{U(1)}]^2 $ \\[2mm]
$L(1,1)$  &  $\frac{F_{4(4)}}{USp(6)\times USp(2)} $
    & $\frac{SU(1,1)}{U(1)}\times
    \frac{Sp(6)}{U(3)} $ \\[2mm]
$L(2,1)$ &  $\frac{E_{6(2)}}{SU(6)\times SU(2)} $& $\frac{SU(1,1)}{U(1)}\times
    \frac{SU(3,3)}{SU(3)\times SU(3)\times U(1)} $ \\[2mm]
$L(4,1)$  &  $\frac{E_{7(-8)}}{SO(12)\times SU(2)} $
    & $\frac{SU(1,1)}{U(1)}\times
    \frac{SO^*(12)}{U(6)} $ \\[2mm]
$L(8,1)$ &  $\frac{E_{8(-24)}}{E_7\times SU(2)} $& $\frac{SU(1,1)}{U(1)}\times
    \frac{E_{7(-26)}}{E_6\times SO(2)} $ \\[2mm] \hline
\end{tabular}
\end{center}
\label{results}
\end{table}

\newpage

\section{Dual K\" ahler spaces}\label{curvature}
In the previous sections we studied the result of the orientifold
truncation (\ref{truncation}) on the symmetric special
quaternionic spaces.  It was noticed in \cite{Ferrara2004} that if
one starts with the very special quaternionic spaces,
characterized by a $d$-symbol $d_{ABC}$, and takes the truncation
$(A) = (\alpha)$, that is $n_{-} = 0$, the resulting K\" ahler
spaces are always of the form $\MM_{trunc} = SU(1,1)/U(1) \times
\KK_{dual}$. When we take orientifold truncations with $n_{-}\neq
0$, no $SU(1,1)/U(1)$ part splits off in general. However, for
specific $d$-symbols\footnote{The fact that in this case again a
$SU(1,1)/U(1)$ part splits off does not depend on the fact that
$d_{\mu ij}$ are $\gamma$-matrices and so there are also many
non-homogeneous VSQ manifolds for which one can find truncations
with $n_{-}\neq 0$ and a $SU(1,1)/U(1)$ part splitting off.}, e.g
those characterizing the HVSQ manifolds, we found a truncation
with $n_{-}\neq 0$ where a $SU(1,1)/U(1)$ part splits off (see
section \ref{splitting}).  We can then, as in \cite{Ferrara2004,
Ferrara0403}, view the remaining manifold as a dual K\" ahler
manifold.  In the previous section, we found that in the symmetric
case, the dual K\" ahler manifolds are the same cosets as the
original very special K\" ahler spaces. In the first subsection,
we will verify that the dual K\" ahler spaces also have the same
curvature tensor in the symmetric case. For the non-symmetric
case, the curvature tensors are different in general, as we will
show by computing the scalar curvature for a specific case.
Another important property to check is whether these dual K\"
ahler spaces give rise to flat potentials, like in
\cite{Ferrara2004}. We will find in the last subsection that this
is indeed the case.
\subsection{The curvature tensor of the truncated  VSQ  \\ manifolds}
We start by looking at the scalar curvature. For the readers
convenience, we have written down the explicit expressions for the
components of the Christoffel connection and curvature tensor in
the appendix.
\\
We restrict ourselves to the homogeneous very special quaternionic
manifolds.  These are characterized by the $d$-symbols
(\ref{dsym}) mentioned in section \ref{trunchvsq}. Inserting these
$d$-symbols in the expressions for the metric (\ref{metric})
\footnote{Remember, we take $(\alpha) = (1,M)$ and $(a)=(i)$.}, it
is easy to see that in this case, the K\" ahler manifolds have an
$\frac{SU(1,1)}{U(1)}$ part that splits off. Indeed, we find $\bar
g^{1M} = \bar g^{1}{}_{a} = \bar g^{1}{}_{0} = 0 $, and therefore
also, since the metric is K\" ahler, that $\bar g^{11}$ only
depends on $t_{1}$ and that all other components are independent
of $t_{1}$. We can now calculate the scalar curvature of the
`dual' K\" ahler space. We have \beqn \bar R &=& \bar
R^{M}{}_{M}{}^{N}{}_{N} + \bar R_a{}^a{}_b{}^b + \bar
R_0{}^0{}_0{}^0  + 2 \bar R^{M}{}_{M}{}_{a}{}^{a} + 2 \bar
R^{M}{}_{M}{}_{0}{}^{0} + 2\bar R_a{}^a{}_0{}^0 , \eeqn with (see
appendix) \beqn && \bar R^{M}{}_{M}{}^{N}{}_{N} = -(q+2)^2 ,
\qquad \bar R_0{}^0{}_0{}^0 = -2  , \qquad \bar
R_{a}{}^{a}{}_{0}{}^{0} = - r , \nonumber \\ && \bar
R_a{}^a{}_b{}^b = - r^2 +
\frac{1}{2}\eta^{MN}\gamma_{Mfa}\gamma_{Nbe}
\frac{(\gamma_{K}^{ae}\chi^{K})(\gamma_{L}^{bf}\chi^{L})}{(\chi^{P}\chi_{P})}
+ \frac{1}{2}\eta^{MN}\gamma_{Mfb}\gamma_{Nae}
\frac{(\gamma_{K}^{ae}\chi^{K})(\gamma_{L}^{bf}\chi^{L})}{(\chi^{P}\chi_{P})} , \nonumber \\
&& \bar R^{M}{}_{M}{}_{a}{}^{a} = -r(q+2) + r -
\frac{1}{2}\eta^{MN}\gamma_{Mfa}\gamma_{Nbe}
\frac{(\gamma_{K}^{ae}\chi^{K})(\gamma_{L}^{bf}\chi^{L})}{(\chi^{P}\chi_{P})}
, \quad
 \bar R^{M}{}_{M}{}_{0}{}^{0} = 0 ,
 \eeqn  where we used the following
expressions \footnote{Here and in the appendix we used a shorthand
notation, $\kappa_{..}\equiv (\kappa \chi)_{..},
(\kappa^{-1})^{..} \equiv ((\kappa \chi)^{-1})^{..}$.} \beqn & &
\kappa = - 3\chi^{M}\chi_{M}\chi^{1} , \quad \kappa_{1} =
-\chi^{M}\chi_{M} , \quad  \kappa_{M} = -2\chi^{1}\chi_{M} ,  \cr & &
\kappa_{11} = 0  , \quad \kappa_{1M} = - \chi_{M}  , \quad \kappa_{MN} =
- \eta_{MN}\chi^{1} , \cr & & \kappa_{a} = 0 , \quad \kappa_{\alpha
a}= 0 , \quad \kappa_{ab} = \gamma_{Mab}\chi^{M} , \cr & &
(\kappa^{-1})^{11} = -\frac{3(\chi^{1})^2}{\kappa} , \quad
(\kappa^{-1})^{1M} = \frac{3 \chi^{M}\chi^{1}}{\kappa} , \cr & &
(\kappa^{-1})^{MN} = - \frac{\eta^{MN}}{\chi^{1}} - \frac{3
\chi^{M}\chi^{N}}{\kappa} , \quad (\kappa^{-1})^{ab} =
\frac{\gamma_{M}^{ab}\chi^{M}}{\chi^{N}\chi_{N}} . \eeqn

Since we know from the previous sections that the K\"ahler
submanifolds we get by doing the above truncation are again
homogeneous, we can calculate the constant scalar curvature in one
convenient point. Thus, we look at a point where $\lambda^{\mu} =
0$, $\lambda^{2} \neq 0$, since the above expressions then
simplify considerably. We get \beqn \bar R = -(q+2)^2 - 2(q+2)r -
r^2 - 2 +
\frac{1}{2}\gamma_{Mfa}\gamma^{M}_{be}\delta^{ae}\delta^{bf} -
\frac{1}{2}\gamma_{Mfb}\gamma^{M}_{ae}\delta^{ae}\delta^{bf} .
\eeqn 
For the homogeneous non-symmetric space $L(0,1,1)$ we find $\bar R = -16$ for the dual space, while the 
original special K\" ahler space has $R = -18$, showing that generically the curvature tensors are different.
\newpage
For the symmetric very special quaternionic manifolds, we
also have $\gamma_{M(fb}\gamma^{M}_{ae)} = 0$, so we finally get
\beqn \bar R = -(q+2)^2 - 2 - \frac{1}{2}q r - r^2 -4r. \eeqn It
is now easy to verify that we get the same value as for the
original special K\"ahler manifolds that gave rise to the
quaternionic manifolds via the \cmap , just like in
\cite{Ferrara0403}. In fact, in the case $r=0$, since there are no
$\gamma_{M}$-matrices, the only difference with the case studied
in \cite{Ferrara2004, Ferrara0403} is the $SU(1,1)$ that was
discarded. The truncated manifold is of the form
$(SU(1,1)/U(1))_{1} \times (SU(1,1)/U(1))_{2} \times \MM$, with
$SU(1,1)_{1}$ generated by $\{ \uu{\hat \beta}^0, \uu{\alpha}_0,
\uu{\epsilon}_0- \frac{3}{2} \uu{\beta} \}$ and $SU(1,1)_{2}$
generated by $\{ \uu{\hat \alpha}_1, \uu{\beta}^1 ,\uu{\epsilon}
\}$.  Taking $\KK_{dual} = (SU(1,1)/U(1))_{2} \times \MM$
corresponds to the case discussed in \cite{Ferrara2004,
Ferrara0403}, while taking $\KK_{dual} = (SU(1,1)/U(1))_{1} \times
\MM$ corresponds to the case discussed in this paper.  Since both
$SU(1,1)/U(1)$ parts have scalar curvature $-2$, as is easy to
verify, there is actually no difference between the two cases and
so the dual K\" ahler spaces and the original very special K\"
ahler spaces have, in the case of symmetric spaces, the same
curvature tensor, like in \cite{Ferrara0403}.
\\
To study the full curvature tensor in the general case where $r
\neq 0$, we make use of an extension of the computer program used
in \cite{Ferrara0403}\footnote{We would like to thank M. Trigiante
for providing us with this program.}.  Let $z^{A}$ denote the
coordinates of the original very special K\" ahler space, and
$y^{A} = \Im z^{A}$.  The metric can then be derived from the K\"
ahler potential $K = - \log(V(y))$ and the curvature tensor is
given by
\begin{equation}
R_A{}^C{}_B{}^D=-2\delta^C_{(A}\delta^D_{B)}+\frac{4}{3}d_{ABE}\,C^{CDE},
\end{equation}
where $C^{ABC}$ is defined as in (\ref{c-up}). In the symmetric
case, it turns out that $C^{ABC}$ is constant \cite{cetal}, so
that the curvature tensor $R_A{}^C{}_B{}^D$ is constant (in this
coordinate system).  Similarly, we will denote the complex
coordinates of the dual K\" ahler space as $(\eta^A) = (\eta,
\epsilon_M, \rho^i)$ (see previous notation) and the curvature
tensor as $\bar R_A{}^C{}_B{}^D$.  All components which are
non-zero in the symmetric case are listed in the appendix and they
can be derived from the K\" ahler potential $K=-2\log V\big(
\chi^\alpha(p(t_\beta,\lambda^a,\eta_1))\big) - \log (\eta_1) $
(see section \ref{kpot}). Using the computer program mentioned
before, we then find that the curvature tensor of the dual spaces
are also constant in the symmetric case. To prove that the
curvature tensors of the original special K\"ahler spaces and
their dual spaces are the same, we need a (linear) coordinate
transformation that relates the two tensors. This transformation
can be found for all cases.  We first transform the dual curvature
tensor $\bar R$ to $\bar R'$,
\begin{equation}\bar
{R'}_A{}^C{}_B{}^D =
U_{A}{}^{A'}U_{B}{}^{B'}R_{A'}{}^{C'}{}_{B'}{}^{D'}(U^{-1})_{C'}{}^{C}(U^{-1})_{D'}{}^{D},
\end{equation}
with $U_{A}{}^{C}(U^{-1})_{C}{}^{B} = \delta_{A}{}^{B}$ and $
U_{A}{}^{B} = v^{A}\delta_{A}^{B}$ where $v^1 = 1$, $v^2 = -2$ ,
$v^{\mu} = 2$ and $v^{i} = 1$, and as before $\mu = 3,..., q+3$
runs over $q+1$ values, and $i = q+4,...,n$ runs over $r$ values.
We then find
\begin{equation}
R_A{}^C{}_B{}^D = \bar{R'}_C{}^A{}_D{}^B.
\end{equation}
In the cases where $r=0$, the transformation $U$ has no effect and
we find $R_A{}^C{}_B{}^D = \bar{R}_C{}^A{}_D{}^B$, as in
\cite{Ferrara0403}, in agreement with our previous arguments.
\subsection{Flat potentials}
We will now show that the dual K\" ahler spaces give rise to flat
potentials in $N=1$ supergravity. In order to establish this, we
need to check the following equality \cite{Ferrara2004}
\begin{equation}
\frac{\partial K}{\partial\Re \eta^A} (\bar
g^{-1})^{AB}\frac{\partial K}{\partial\Re \eta^B}=3 ,
\end{equation}
where $(\eta^A) = (\eta, \epsilon_M, \rho^i)$. A straightforward
calculation, using equations (\ref{kahlerpot}),(\ref{invmetric})
and (\ref{dsym}), shows that this is indeed the case. Notice that
$\frac{\partial K}{\partial t_{M}}$, $\frac{\partial K}{\partial
\lambda^i}$ and $\frac{\partial K}{\partial \eta_{1}}$ are
independent of $t_{1}$ since $\bar g^{1M} = \bar g ^{1}{}_{i} =
\bar g^{1}{}_{0} = 0$. The calculations do not depend on the
specific form of e.g $d_{\mu ij}$ and can therefore be generalized
to non-homogeneous K\" ahler manifolds. Thus, we have found a new
class of K\" ahler potentials that lead to flat potentials. \\
Notice that all the results in this section depend on the specific
form (\ref{dsym}) of the d-symbols and so do not apply to the case
$SG_{5}$, characterized by e.g. $d_{111} = 1$. In this case the
original special K\" ahler space is $SU(1,1)/U(1)$ with scalar
curvature $-2/3$ and $\MM_{trunc} = (SU(1,1)/U(1))^2$, with one
$SU(1,1)/U(1)$ also having scalar curvature $-2/3$ and the other
having scalar curvature $-2$.  The $SU(1,1)/U(1)$ manifold with
scalar curvature $-2/3$ gives rise to a flat potential, the other
one does not.
\section{Conclusion}
In this paper, we studied the orientifold truncation that arises
when compactifying type II string theory on Calabi-Yau
orientifolds with $O3/O7$-planes. This truncation provides a way
of reducing supersymmetry from $N=2$ to $N=1$, and we have studied
the reduction of the hypermultiplet sector of $N=2$ supergravity
with scalars spanning a ($4n+4$)-dimensional very special
quaternionic-K\"ahler manifold. We have established that the
truncated ($2n+2$)-dimensional manifold is K\"ahler, by explicitly
determining the K\"ahler potential. Further, we studied the
truncated isometry algebra. For homogeneous very special
quaternionic-K\"ahler spaces, in the generic case, the truncated
manifold splits into $SU(1,1)/U(1)$ times a $2n$-dimensional
K\"ahler space that can be viewed as dual to the original special
K\"ahler space, in the sense of \cite{Ferrara2004}. These spaces
also give rise to flat potentials. For the case of symmetric
spaces, the dual K\"ahler space consists of the same coset space
as the original special K\"ahler space and has the same curvature
tensor. We also studied a more general truncation for the
symmetric spaces $L(0,P)$, and found that generically no
$SU(1,1)/U(1)$ splits off anymore. Finally, one can also truncate
symmetric special quaternionic-K\"ahler spaces that are not very
special (this includes the case of minimal coupling): in this case
the truncated manifold is still symmetric, but does
not always factorize. The results are summarized in table~\eqn{results}. \\

\noindent{ \bf Acknowledgements}\vspace{0.3cm}

We thank A. Van Proeyen for many helpful discussions, and for proofreading the manu\-script.
We have also greatly benefited from correspondence with S. Ferrara, R. D'Auria, M. Trigiante and T.W. Grimm.
This work was supported in part by the European Community's Human Potential Programme
under contract HPRN-CT-2000-00131 Quantum Spacetime and by the Belgian Federal Office for Scientific,
Technical and Cultural Affairs through the Inter-University Attraction Pole P5/27.

\appendix
\section{Appendix: calculation of the curvature tensor.}
In this appendix we will give the components of the Christoffel
connection and the curvature tensor which are necessary to
calculate the scalar curvature of the truncated very special
quaternionic-K\" ahler manifold. We also give all other components
of the curvature tensor which are non-zero in the symmetric case.
The connection and the Riemann tensor are defined as \beqn \bar
\Gamma^{\Lambda}{}_{\Sigma \Delta} &=& (\bar g^{-1})^{\Lambda \bar
\Lambda}\partial_{\Sigma}
\bar g_{\bar \Lambda \Delta}, \nonumber \\
\bar R_{\Sigma}{}^{\Gamma}{}_{\Delta}{}^{\Lambda} &=& - (\bar
g^{-1})^{\Gamma \bar \Gamma}\partial_{\bar \Gamma}\bar
\Gamma^{\Lambda}{}_{\Sigma \Delta}.\eeqn  Because the metric $\bar
g_{\Lambda \bar \Sigma}$ is K\" ahler, the Christoffel connection
is symmetric in the last two indices and the curvature tensor is
symmetric under exchange of the first and the third index and/or
the second and the fourth index. \\
After some long calculations, using equations (\ref{metric}) and
(\ref{invmetric}) in section \ref{kpot}, we found the following
results \footnote{We used the following shorthand notation,
$\kappa_{..}\equiv (\kappa \chi)_{..}, (\kappa^{-1})^{..} \equiv
((\kappa \chi)^{-1})^{..}$.} \beqn \bar \Gamma_{\gamma}{}^{\alpha
\beta} &=& (\bar g^{-1})_{\gamma \bar \gamma}\frac{\partial \bar
g^{\bar \gamma \beta}}{\partial t_{\alpha}} + (\bar
g^{-1})_{\gamma}{}^{0} \frac{\partial \bar
g_{0}{}^{\beta}}{\partial t_{\alpha}} + (\bar
g^{-1})_{\gamma}{}^{a}
\frac{\partial \bar g_{a}{}^{\beta}}{\partial t_{\alpha}} \nonumber \\
&=& d_{\bar \alpha \bar \beta \gamma}(\kappa^{-1})^{\alpha \bar
\alpha} (\kappa^{-1})^{\beta \bar \beta} +
\frac{6}{\kappa}\delta_{\gamma}^{(\alpha}\lambda^{\beta)} -
\frac{3}{\kappa} (\kappa^{-1})^{\alpha
\beta}\kappa_{\gamma} , \nonumber\\
\bar \Gamma_{\gamma}{}^{\alpha}{}_0 &=&
\frac{1}{2\eta_{1}^2}(\kappa \lambda \lambda)_{\beta}\bar
\Gamma_{\gamma}{}^{\alpha \beta} + \frac{1}{\eta_{1}^2}(\kappa
\lambda)_{a \bar \gamma}(\kappa^{-1})^{\bar \gamma
\alpha}(\kappa^{-1})^{a c}(\kappa \lambda)_{\gamma c} -
\frac{3}{\kappa \eta_{1}^2}(\kappa \lambda
\lambda)_{\gamma}\lambda^{\alpha} , \nonumber\\
\bar\Gamma_{\gamma}{}^{\alpha}{}_a &=& -\frac{1}{\eta_{1}}(\kappa
\lambda \lambda)_{\beta a}\bar \Gamma_{\gamma}{}^{\alpha \beta} -
\frac{1}{\eta_{1}}d_{a b \bar \gamma}(\kappa^{-1})^{\bar \gamma
\alpha}(\kappa^{-1})^{b c}(\kappa \lambda)_{\gamma c} +
\frac{3}{\kappa \eta_{1}}(\kappa \lambda)_{\gamma a}
\lambda^{\alpha} , \nonumber\\
\bar \Gamma^{a}{}^{\alpha}{}_b &=& -(\kappa^{-1})^{a c}d_{b  c
\bar \gamma}(\kappa^{-1})^{\bar \gamma \alpha} +
\frac{3}{\kappa}\delta^{a}_{b}\lambda^{\alpha} , \, \qquad \bar
\Gamma^{0}{}^{\alpha}{}_0 = 0 , \, \qquad
\bar \Gamma^{a}{}^{\alpha}{}_0 = - \bar \Gamma^{a}{}^{\alpha}{}_b \frac{\lambda^{b}}{\eta_{1}} ,\nonumber\\
\bar \Gamma^{0}{}_{a 0} &=&  \frac{1}{\eta_{1}}G_{a b}\lambda^{b},
\, \qquad \bar \Gamma^{0}{}_{0 0} = -\frac{1}{\eta_{1}^2}G_{a
b}\lambda^{a}\lambda^{b} - \frac{2}{\eta_{1}} , \, \qquad \bar \Gamma^{0}{}_{ab} = - G_{ab} ,\nonumber\\
\bar \Gamma^{a}{}_{b 0} &=& \frac{1}{2 \eta_{1}^2}d_{b c \bar
\gamma}(\kappa^{-1})^{\bar \gamma \alpha}(\kappa \lambda
\lambda)_{\alpha}(\kappa^{-1})^{ac} -
\frac{1}{\eta_{1}^2}(\kappa^{-1})^{a c}(\kappa \lambda)_{\bar
\gamma  c}(\kappa \lambda)_{\bar \beta b}
(\kappa^{-1})^{\bar \gamma \bar \beta}  \nonumber \\
& &  + \frac{3}{\kappa \eta_{1}^2}(\kappa \lambda \lambda)_{b}
\lambda^{a} + \frac{3}{2 \kappa \eta_{1}^2} (\kappa \lambda
\lambda)_{\alpha}\lambda^{\alpha} \delta^{a}_{b} - \frac{6
\lambda^a}{\kappa \eta_{1}^2}\kappa_{b c} \lambda^c -
\frac{1}{\eta_{1}}\delta^{a}_{b} , \nonumber \\ \bar
\Gamma^{a}{}_{b c} &=&  \frac{2}{\eta_{1}}d_{\bar \gamma d
(c}(\kappa \lambda)_{b) \alpha}(\kappa^{-1})^{\bar \gamma
\alpha}(\kappa^{-1})^{a d} - \frac{6}{\kappa
\eta_{1}}\delta^{a}_{(b}(\kappa \lambda \lambda)_{c)} + \frac{6
\lambda^a}{\kappa \eta_{1}}\kappa_{b c} , \nonumber \eeqn \beqn
\bar \Gamma_{\alpha}{}_{ab} &=&
 \frac{2}{\eta_{1}^2}(\kappa^{-1})^{ce}(\kappa^{-1})^{\beta
\bar \beta}(\kappa \lambda)_{\alpha e}d_{\beta c (a}(\kappa
\lambda)_{b)\bar \beta} - \frac{3}{\kappa
\eta_{1}^2}(\kappa^{-1})^{\beta \gamma}(\kappa \lambda)_{\beta
a}(\kappa \lambda)_{\gamma b}\kappa_{\alpha}\nonumber \\ & & + \frac{d_{\bar \beta \bar \gamma
\alpha}}{\eta_{1}^2}(\kappa^{-1})^{\beta \bar
\beta}(\kappa^{-1})^{\gamma \bar \gamma}(\kappa \lambda)_{\beta
a}(\kappa \lambda)_{\gamma b} - \frac{d_{\alpha a b}}{\eta{1}} +
\frac{3}{\kappa \eta_{1}^2}(\kappa \lambda
\lambda)_{\alpha}\kappa_{a b}, \eeqn

\beqn \bar R^{\beta}{}_{\alpha}{}^{\mu}{}_{\gamma} &=& -(\bar
g^{-1})_{\alpha \bar \alpha}\,\frac{\partial}{\partial t_{\bar
\alpha}}\,\bar \Gamma_{\gamma}{}^{\beta \mu } -(\bar
g^{-1})_{\alpha}{}^{a}\,\frac{\partial}{\partial
\lambda^{a}}\,\bar \Gamma_{\gamma}{}^{\beta \mu }-(\bar
g^{-1})_{\alpha}{}^{0}\,\frac{\partial}{\partial
\eta_{1}}\,\bar \Gamma_{\gamma}{}^{\beta \mu }\nonumber\\
& = &-\delta^{\beta}_{(\alpha}\delta^{\mu}_{\gamma)}-d_{\bar \beta
\bar \mu(\alpha}\,\kappa_{\gamma)}(\kappa^{-1})^{\beta \bar
\beta}\,
(\kappa^{-1})^{\mu \bar \mu}+(\kappa^{-1})^{\beta \mu}\,\kappa_{\alpha \gamma} \nonumber \\
& & +\frac{\kappa}{3}\,(\kappa^{-1})^{\bar \beta
(\beta}(\kappa^{-1})^{\mu) \bar \mu}(\kappa^{-1})^{\delta \bar
\delta}d_{\bar \beta \delta \alpha}\,d_{\bar \mu \bar \delta
\gamma} , \nonumber\\
\bar R_c{}^a{}_b{}^d&=&-(\bar g^{-1})^{a}{}_{\bar
\alpha}\,\frac{\partial}{\partial t_{\bar \alpha}
}\,\bar\Gamma^{d}{}_{bc}-(\bar g^{-1})^{a
e}\,\frac{\partial}{\partial \lambda^{e}
}\,\bar\Gamma^{d}{}_{bc}-(\bar g^{-1})^{a
0}\,\frac{\partial}{\partial
\eta_{1}}\,\bar\Gamma^{d}{}_{bc}\nonumber\\
&=&-\delta^a_{(c}\delta^d_{b)}+(\kappa^{-1})^{ad}\,\kappa_{cb}+\frac{\kappa}{3}\,
(\kappa^{-1})^{e(a}(\kappa^{-1})^{d)f}(\kappa^{-1})^{\gamma \bar
\gamma}d_{\gamma ec}\,d_{\bar \gamma fb}
 , \nonumber\\
\bar R_0{}^0{}_0{}^0 &=& -2 , \, \qquad  \bar
R^{\beta}{}_{\alpha}{}_{0}{}^{0}  =  0 , \, \qquad
\bar R_{b}{}^{a}{}_{0}{}^{0}  =  - \delta^{a}_{b} , \, \qquad R_{a}{}_{\gamma}{}_{b}{}^{0}  =  - d_{\gamma a b},\nonumber\\
\bar R^{\beta}{}_{\alpha}{}_{b}{}^{a} & = &-\frac{1}{2}\delta^{\beta}_{\alpha}\delta^{a}_{b}+ \frac{1}{2}d_{\bar \beta e b}\,\kappa_{\alpha}(\kappa^{-1})^{\beta \bar \beta}(\kappa^{-1})^{a e} -\frac{\kappa}{6}\,(\kappa^{-1})^{a e}(\kappa^{-1})^{c f}(\kappa^{-1})^{\bar \gamma \beta}d_{\alpha c e }\,d_{\bar \gamma b f}\nonumber \\
& & -\frac{\kappa}{6}\,(\kappa^{-1})^{a e}(\kappa^{-1})^{\beta
\gamma}(\kappa^{-1})^{\bar \gamma \bar \beta}d_{\alpha \gamma \bar
\gamma}\,d_{\bar \beta  b e},\nonumber\\
\bar R^{\alpha a}{}_{0}{}^{b} & = &
\frac{\kappa}{6}\,(\kappa^{-1})^{a e}(\kappa^{-1})^{b
c}(\kappa^{-1})^{\gamma \alpha}d_{e c \gamma} -
\frac{1}{2}(\kappa^{-1})^{a b} \lambda^{\alpha} .\eeqn All other
components of the curvature tensor are zero in points of the
manifold were $\lambda^{a} = 0$ for all $a$.  For the dual
symmetric spaces this implies that these components are zero in
every point.

\end{document}